\RequirePackage{fix-cm}
\documentclass[twocolumn,epjc3]{svjour3}
\usepackage{graphicx,amsmath,amssymb,booktabs}
\usepackage[numbers,sort&compress]{natbib}
\PassOptionsToPackage{colorlinks=true,linkcolor=blue,citecolor=red,urlcolor=magenta}{hyperref}
\usepackage{orcidlink}
\numberwithin{equation}{section}
\smartqed
\journalname{Eur. Phys. J. C}
\hypersetup{
  pdftitle={Revisiting the holographic thermodynamics of charged Gauss--Bonnet AdS black holes},
  pdfauthor={Limin Zeng}
}
\newcommand{\dd}{\mathrm d}
\newcommand{\VV}{\mathcal V}
\newcommand{\Gh}{\mathcal G}
\newcommand{\Tb}{T_{\mathrm b}}
\newcommand{\aold}{\widetilde\alpha}
\newcommand{\Aold}{\widetilde{\mathcal A}}
\newcommand{\Tcoex}{T_{\mathrm{b,coex}}}
\newcommand{\Tbc}{T_{\mathrm{b,c}}}

\begin{document}
\title{Revisiting the holographic thermodynamics of charged Gauss--Bonnet AdS black holes}

\author{Limin Zeng\orcidlink{0009-0003-5703-2687}\thanksref{e1,addr1,addr2,addr3}}
\thankstext{e1}{e-mail: zenglimin25@mails.ucas.ac.cn}
\institute{School of Fundamental Physics and Mathematical Sciences,\\
  Hangzhou Institute for Advanced Study, UCAS, Hangzhou 310024, China \label{addr1}
  \and
  Institute of Theoretical Physics, Chinese Academy of Sciences, Beijing 100190, China \label{addr2}
  \and
  University of Chinese Academy of Sciences, Beijing 100049, China \label{addr3}}
\date{}
\maketitle

\begin{abstract}
We revisit the holographic thermodynamics of charged spherical Gauss--Bonnet AdS black holes. We minimize the free energy at the prescribed bath temperature, minus the external source times the curvature coupling, over horizon radius and coupling. In four bulk dimensions the logarithmic entropy makes this potential strictly convex in the coupling at positive bath temperature. When the minimizing coupling lies inside the allowed interval, its adjustment lowers the radial curvature and explains the two temperature turning points in the reference example. Two distinct radial minima can exchange preference at equal potential. In the higher-charge example a single preferred minimum reaches a coupling bound continuously. Numerical continuation at the three sources studied shows that the coexisting minima leave their coupling bounds in succession and merge at an interior critical point. A quartic expansion gives square-root closure of the radius, coupling, and entropy gaps. At the reference source it also gives a heat-capacity divergence with exponent two thirds on a temperature sweep at the critical charge and as the inverse distance in charge along coexistence. In five dimensions we give a fixed-coupling benchmark and compare the two coupling bounds at fixed source under two bath conditions, one using time normalized by the bare AdS length and the other using proper time on the boundary cylinder.

\end{abstract}

\section{Introduction}
\label{sec:intro}

Charged spherical AdS black holes can have several regular radii at the same boundary temperature. Their canonical free energies can form a swallowtail, with the crossing giving small--large black-hole coexistence \cite{chamblin}. In Gauss--Bonnet gravity the curvature coupling modifies the temperature and entropy and may itself enter the first law as a thermodynamic variable \cite{cai,kastorLovelockFirstLaw,caiPV,gbcoupling,wei}. For four-dimensional charged Gauss--Bonnet black holes, the free energy as a function of horizon radius has also been used to identify coexistence from two minima of equal depth \cite{liWangLandscape4d}. When the coupling varies along a plotted curve, the projected folds and crossings must be interpreted using the work terms for that path.

Early work on holographic black hole chemistry related bulk scaling to the number of degrees of freedom of the boundary theory \cite{karchRobinson}. Subsequent formulations clarified the chemical potential for the number of degrees of freedom and the role of the boundary conformal factor \cite{congC,visserColor,ahmedHolographicFirstLaw,congCFT}. In Gauss--Bonnet gravity, related work has investigated how the curvature coupling affects central-charge criticality \cite{kumarGaussBonnetCriticality,quGaussBonnetCriticality}. Higher-derivative bulk--boundary first laws can also be formulated using two independent boundary central charges \cite{puniaHigherDerivative,cuiFanCentralCharges}.

Earlier holographic studies have examined phase and response diagrams of charged Gauss--Bonnet black holes \cite{original,epjc}. Here we fix an external source \(J\) conjugate to the dimensionless curvature coupling \(\lambda\). We compare black holes using \(E-T_{\rm b}S-J\lambda\), where \(E\) is the energy, \(S\) the entropy, and \(T_{\rm b}\) the imposed bath temperature. The coupling may adjust within a specified interval and can differ between the competing black holes. We ask how this adjustment affects the preferred black hole as an external control varies. In four bulk dimensions the coupling that minimizes this potential at each radius is available in closed form. When it lies inside the allowed interval, its adjustment lowers the radial curvature and explains the turning points in the temperature of the stationary black holes. The radial comparison distinguishes an equal-potential exchange of two minima from a continuous contact of one minimum with a coupling bound. We follow the equal-potential minima numerically in charge to locate the bound departures and the critical endpoint. The same potential gives the pressure and electric-potential jumps when volume or charge is scanned.

Our four-dimensional comparison uses the charged spherical branch obtained from the higher-dimensional Gauss--Bonnet solution by rescaling the coupling before taking the four-dimensional limit \cite{fernandesCharged4d,epjc}. We use the surface-gravity temperature. The radial first law gives the logarithmic dependence of the entropy, and we specify its reference length. The boundary temperature is measured in time normalized using the bare AdS length, which we call the bare-length clock, as in Ref.~\cite{epjc}. We use background subtraction to fix the energy zero. Section~\ref{sec:boundary-conventions} defines this time convention and relates it to the proper time of a static observer on the boundary sphere, which defines the cylinder clock used for the five-dimensional comparison in Ref.~\cite{across}. \ref{app:five-source} uses the same fixed-source minimization as in four dimensions under both bath conditions. The horizon radius and the coupling can both vary. In five bulk dimensions, changing the coupling changes the boundary Weyl anomaly coefficients, so this comparison spans a family of CFTs \cite{buchelMyers,across}.

The paper is organized as follows. Section~\ref{sec:method} derives the boundary differential and the equal-potential criteria. Section~\ref{sec:four} solves the four-dimensional fixed-source problem and follows the coexistence line to its critical endpoint. Section~\ref{sec:five} gives a five-dimensional fixed-coupling benchmark, and Sec.~\ref{sec:responses} tests the response diagrams. \ref{app:formulas} derives the state functions, the first law in a dimensionful coupling coordinate, and the radial temperature derivatives. \ref{app:numerics} specifies the admissible domain and documents the checks of coexistence, critical scaling, and the response diagrams. \ref{app:five-source} compares coexistence between the two coupling bounds in five dimensions at fixed source for two choices of bath. Each comparison fixes a common temperature in its own clock, either bare-length or proper cylinder time.

\section{Boundary variables and path criteria}
\label{sec:method}

\subsection{Conventions and first law}
\label{sec:boundary-conventions}

Let \(d\) denote the bulk dimension. The starting point for \(d\geq5\) is the Einstein--Maxwell--Gauss--Bonnet bulk action with curvature coupling \(\alpha_{\rm GB}\) \cite{cai,epjc},
\begin{multline*}
 I_{\rm bulk}=\frac{1}{16\pi G}\int \dd^d x\,\sqrt{-g}\\
 \times\left(\mathcal R-2\Lambda+\alpha_{\rm GB}\mathcal L_{\rm GB}\right)
 +I_{\rm Maxwell}.
\end{multline*}
Here \(G\) is Newton's constant, \(\Lambda=-(d-1)(d-2)/(2L^2)\) fixes the bare AdS length \(L\), and \(I_{\rm Maxwell}\) is the bulk Maxwell action. The curvature density is
\begin{equation*}
 \mathcal L_{\rm GB}=\mathcal R_{\mu\nu\rho\sigma}\mathcal R^{\mu\nu\rho\sigma}
 -4\mathcal R_{\mu\nu}\mathcal R^{\mu\nu}+\mathcal R^2 .
\end{equation*}
The charged spherical equation \eqref{eq:metric} contains the combination \(\alpha=(d-3)(d-4)\alpha_{\rm GB}\). We take the four-dimensional limit of this equation at fixed \(\alpha\), which amounts to rescaling the action coupling before taking the limit. We use the root continuous with the Einstein--Maxwell solution at zero coupling \cite{fernandesCharged4d,wei}. Let \(r_+\) be the outer horizon radius, \(Q_b\) the bulk electric charge, and \(R\) the boundary sphere radius. We normalize \(Q_b\) as specified below Eq.~\eqref{eq:horizon-mass}. The boundary variables used below are
\begin{align}
 x&=r_+/L,\qquad \lambda=\alpha/L^2,\qquad \VV=R^{d-2},\\
 C&=\frac{L^{d-2}}{16\pi G},\qquad
 Q=\frac{Q_bL}{\sqrt G},\qquad q=\frac{Q}{16\pi C}.
 \label{eq:variables}
\end{align}
We absorb the unit-sphere volume into the normalization of the state functions. The parameter \(C\) sets the overall boundary normalization, and \(Q\) is the boundary charge. At fixed \(C\), changes in \(\lambda\) can change the holographic theory data. In five bulk dimensions they change the boundary Weyl anomaly coefficients \cite{buchelMyers,across}. The main text denotes boundary thermodynamic quantities by \(E,T,S,Q,p,\Phi,\mu,\Xi\). \ref{app:formulas} uses tildes for the responses and coupling coordinate of the dimensionful convention.

Use the bulk time \(t_{\rm bulk}\) in which the metric has time coefficient \(-f(r)\) and let \(T_H\) be its Hawking temperature. We follow the normalization of Ref.~\cite{epjc} by using \(t_{\rm bare}=(R/L)t_{\rm bulk}\). This gives the state temperature \(T=(L/R)T_H\) and the energy \(E=(L/R)M\), where \(M\) is measured relative to the AdS vacuum at the same coupling. We use \(\Tb\) for the prescribed bath temperature in this clock. Its coexistence and critical values are \(\Tcoex\) and \(\Tbc\). A regular thermal saddle has \(T=\Tb\).

To relate this convention to the proper time of a boundary observer, rescale the bulk metric by \(R^2/r^2\) and take the boundary limit. The angular term is then \(R^2\dd\Omega_{d-2}^2\). On the Einstein branch the asymptotic form is \(f(r)\sim r^2/L_{\rm eff}^2\), where the effective AdS length \(L_{\rm eff}\) follows from the vacuum limit of Eq.~\eqref{eq:metric}. The temporal term in the bare clock is therefore \(-(L/L_{\rm eff})^2\dd t_{\rm bare}^2\). Unit lapse on the same sphere is obtained with \(t_{\rm cyl}=(R/L_{\rm eff})t_{\rm bulk}\). In particular,
\begin{equation}
 T_{\rm cyl}=\beta(\lambda) T,\qquad
 \beta(\lambda)=\frac{L_{\rm eff}}{L}
      =\sqrt{\frac{1+\sqrt{1-4\lambda}}{2}} .
 \label{eq:clock-conversion}
\end{equation}
Let \(T_{\mathrm{b,cyl}}\) denote the imposed bath temperature measured in proper time on the boundary cylinder. The factor \(\beta(\lambda)\) converts the bare-clock bath temperature and background-subtracted energy to the cylinder convention. A common bare-clock bath thus corresponds to different proper bath temperatures across couplings. To implement the common proper bath of Ref.~\cite{across} with the bare-clock equations, set \(\Tb=T_{\mathrm{b,cyl}}/\beta(\lambda)\) separately at each coupling to find its thermal saddles. Evaluate their potentials using the cylinder energy and source defined in \ref{app:five-source}.

In four dimensions the radial first law with the surface-gravity temperature fixes the logarithmic dependence of the entropy on the horizon radius. We choose \(\sqrt{\alpha}\) as the reference length of the logarithm. \ref{app:formulas} gives the derivation. In five dimensions we use the Wald entropy. The state functions are
\begin{align}
 E_4&=\frac{8\pi C}{R}\left(x^3+x+\frac{\lambda+q^2}{x}\right),\label{eq:e4}\\
 S_4&=16\pi^2 C\left[x^2+4\lambda\log(x/\sqrt{\lambda})\right],\label{eq:s4}\\
 RT_4&=\frac{3x^4+x^2-\lambda-q^2}{4\pi x(x^2+2\lambda)}, \label{eq:t4}\\[2pt]
 E_5&=\frac{C}{R}\left[3(x^4+x^2+\lambda)+\frac{q^2}{x^2}\right],\label{eq:e5}\\
 S_5&=4\pi C(x^3+6\lambda x),\label{eq:s5}\\
 RT_5&=\frac{6x^6+3x^4-q^2}{6\pi x^3(x^2+2\lambda)}.
 \label{eq:t5}
\end{align}
\ref{app:formulas} gives the horizon derivation and the entropy normalization. Equations~\eqref{eq:e4}--\eqref{eq:t4} specify the spherical thermodynamics used in all four-dimensional comparisons below.

At fixed \(\lambda,Q,C,R\), both sets of state functions satisfy \(E_x=TS_x\). When the coupling changes at fixed entropy, the radius changes by \(-S_\lambda\,\dd\lambda/S_x\). This yields the fixed-entropy coupling response
\begin{align}
 \Xi&=\left(\frac{\partial E}{\partial\lambda}\right)_{S,Q,C,R}
     =E_\lambda-TS_\lambda,\\
 \Xi_4&=\frac{8\pi C}{Rx}
        -64\pi^2 CT_4\left[\log(x/\sqrt\lambda)-\tfrac12\right],\\
 \Xi_5&=\frac{3C}{R}-24\pi CxT_5.
 \label{eq:xi}
\end{align}
The derivatives on the right of the first line hold \(x\) fixed. The pressure follows from the \(R^{-1}\) scaling of \(E\), since \(S\) is independent of \(R\) in these variables. At fixed \(x,\lambda,R\), simultaneous rescaling of \(Q\) and \(C\) scales \(E\) and \(S\) linearly. The resulting first law and responses are
\begin{align}
 \dd E&=T\,\dd S-p\,\dd\VV+\Phi\,\dd Q+\mu\,\dd C+\Xi\,\dd\lambda,
 \label{eq:firstlaw}\\
 p&=\frac{E}{(d-2)\VV},\qquad C\mu=E-TS-\Phi Q,\nonumber\\
 \Phi_4&=\frac{q}{Rx},\qquad \Phi_5=\frac{q}{8\pi Rx^2}.
 \label{eq:responses}
\end{align}
In particular the entropy correction to \(\mu\) accounts for the radius change needed when \(C\) varies at fixed \(S\). \ref{app:formulas} gives the first law in a dimensionful coupling coordinate. That relation retains an additional term when \(G\) varies.

\subsection{Potentials and equal-area conditions}
\label{sec:potentials}

The Helmholtz potential \(F=E-TS\) obeys
\begin{equation}
 \dd F=-S\,\dd T-p\,\dd\VV+\Phi\,\dd Q+\mu\,\dd C+\Xi\,\dd\lambda.
 \label{eq:dF}
\end{equation}
On a fixed-coupling slice the last term vanishes. When a plot changes the coupling, its slope includes the corresponding work. The state response \(\Xi\) is given by Eq.~\eqref{eq:xi}. The external source \(J\) is a prescribed number, common to every saddle in a comparison. It has units of energy because \(\lambda\) is dimensionless. It enters the potential as \(-J\lambda\), so increasing the source lowers the potential of a larger-coupling saddle relative to that of a smaller-coupling saddle. On a thermal saddle at \(\Tb\) the resulting potential is \(F-J\lambda\), with differential
\begin{equation}
 \dd(F-J\lambda)=\dd F-J\,\dd\lambda-\lambda\,\dd J.
 \label{eq:dFJ}
\end{equation}
On a thermal saddle that minimizes the potential at an interior coupling, the response equals the external source, \(\Xi=J\). If a fixed endpoint of the prescribed coupling interval is active, the derivative \(\Xi-J\) is nonnegative at the lower endpoint and nonpositive at the upper one. Along a branch whose minimizing coupling stays at a fixed bound (a pinned branch), \(\dd\lambda=0\). The coupling work in \(\dd(F-J\lambda)\) therefore vanishes along a regular interior or pinned branch at fixed \(J\).

To compare two endpoints, integrate the relevant potential differential along a continuous path of black-hole states between them. For a mechanical comparison, the potential is \(F+p_0\VV\) at a common imposed pressure \(p_0\). The path varies \(\VV\) at fixed \(T=\Tb\), \(Q\), and \(C\). For an electrical comparison, the potential is \(F-\Phi_0 Q\) at a common imposed electric potential \(\Phi_0\). The path varies \(Q\) at fixed \(T=\Tb\), \(\VV\), and \(C\). On these paths the responses \(p\) and \(\Phi\) may vary while the imposed values \(p_0\) and \(\Phi_0\) in the potentials are constant.

For a thermal comparison, the potential is \(E-\Tb S\) at a common imposed bath temperature \(\Tb\). The path varies \(S\) at fixed \(Q,\VV,C\). Its state temperature \(T\) can vary between the endpoints, which both have \(T=\Tb\). In all three cases the coupling may vary along the path. Equality of the endpoint potentials gives
\begin{align}
 \int_1^2(p-p_0)\,\dd\VV&=\int_1^2\Xi\,\dd\lambda,
       \label{eq:area-p}\\
 \int_1^2(T-\Tb)\,\dd S&=-\int_1^2\Xi\,\dd\lambda,
       \label{eq:area-t}\\
 \int_1^2(\Phi-\Phi_0)\,\dd Q&=-\int_1^2\Xi\,\dd\lambda .
       \label{eq:area-q}
\end{align}
At fixed coupling the right-hand sides vanish, giving the signed equal-area conditions. At fixed source we subtract \(J\lambda\) from each comparison potential and replace \(\Xi\) on the right-hand sides by \(\Xi-J\). If further controls vary along the chosen path, Eq.~\eqref{eq:firstlaw} supplies their work terms. In a numerical scan with disconnected branches, the endpoint potentials can be calculated directly. The integral offers a check when a continuous comparison path is available. An equilibrium assignment also requires local stability and comparison with the other admissible states.

A preferred state is an admissible spherical black-hole candidate that attains the lowest value of the specified classical potential. We call that minimum value the preferred potential. Coexistence means that two distinct candidates attain this minimum at the same controls. The comparison is over a family of couplings and uses the classical potentials themselves.

\section{Four-dimensional fixed-source equilibria}
\label{sec:four}

\subsection{Analytic elimination of the coupling}

We fix the boundary charge \(Q\), normalization \(C\), sphere radius \(R\), bath temperature \(\Tb\), and external source \(J\). The radius \(x\) and the coupling \(\lambda\) can vary. The fixed-source potential is
\begin{equation}
 \Gh_J(x,\lambda)=E(x,\lambda)-\Tb S(x,\lambda)-J\lambda .
 \label{eq:sourcepot}
\end{equation}
The term \(\Tb S\) uses the prescribed bath temperature at every trial point, including those where \(T\ne\Tb\). First prescribe \(0<\lambda_-\leq\lambda\leq\lambda_+<1/4\). At each radius we also require an outer horizon with positive state temperature and positive entropy. We include their limiting values when testing the potential on the feasibility boundary. These conditions can lower the upper bound at that radius to \(\lambda_{\rm cap}(x)\), given by Eq.~\eqref{eq:feasible-coupling-cap}. The reference length in Eq.~\eqref{eq:s4} enters both the potential and the positive-entropy condition. Together with the specified source, it fixes the comparison between states at different couplings.

At fixed \(x\), the four-dimensional energy and the source term are affine in \(\lambda\). The logarithmic entropy then gives the entire coupling curvature \(32\pi^2 C\Tb/\lambda>0\) at positive bath temperature. The unconstrained positive-coupling solution of \(\partial_\lambda\Gh_J=0\) is
\begin{equation}
 \lambda_*(x)=x^2\exp\!\left[
     -1-\frac{8\pi C/(Rx)-J}{32\pi^2 C\Tb}\right].
 \label{eq:lambda}
\end{equation}
At each feasible radius the minimizing coupling \(\lambda_J\) is \(\lambda_*\) clipped to \([\lambda_-,\lambda_{\rm cap}(x)]\). Inserting this value defines the reduced potential \(\widehat{\Gh}_J(x)=\Gh_J(x,\lambda_J)\) at the specified controls. We compare its radial minima and feasibility boundaries to find the preferred state. Where the coupling minimum is smooth and interior, the reduced radial curvature is
\begin{equation}
 \widehat{\Gh}_J''=\Gh_{J,xx}
                  -\frac{\Gh_{J,x\lambda}^2}{\Gh_{J,\lambda\lambda}},
 \label{eq:reduced}
\end{equation}
obtained by differentiating the coupling stationarity equation. On a smooth branch pinned to a fixed endpoint the curvature is \(\Gh_{J,xx}\).

\begin{figure}[t]
\centering
\includegraphics[width=\linewidth]{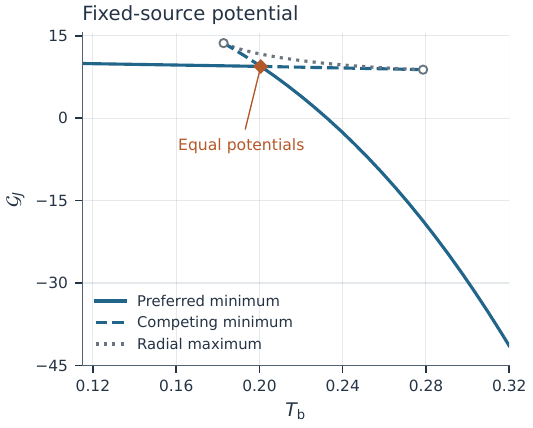}
\caption{Fixed-source potential \eqref{eq:sourcepot} on the stationary branches of Eq.~\eqref{eq:radial-fixed-source} for \(Q=10,C=2,R=1,J=0.005\) and \(\lambda\in[0.001,0.08]\). The curvature in Eq.~\eqref{eq:reduced} distinguishes the radial extrema. Solid, dashed, and dotted lines denote preferred minima, competing local minima, and radial maxima. Circles mark the folds of Eq.~\eqref{eq:fold-coupling-response}. At the diamond the two minima have equal potential and exchange preference. Figure~\ref{fig:diagnostic}(b) shows the coupling profile around the folds and the coupling-bound contacts.}
\label{fig:four}
\end{figure}

We construct the stationary profile by varying the radius and solving the radial stationarity condition, Eq.~\eqref{eq:radial-fixed-source}, for the bath temperature at each radius. Figure~\ref{fig:diagnostic}(a,b) shows this temperature and its minimizing coupling as functions of radius. Figure~\ref{fig:four} instead plots the potential against bath temperature along the same stationary profile. A coupling-bound contact is a point on the profile where the minimizing coupling reaches or leaves the prescribed lower or upper bound. We use the one-sided derivatives there.

\subsection{Temperature folds and contacts with coupling bounds}

\begin{figure*}[t]
\centering
\includegraphics[width=.97\linewidth]{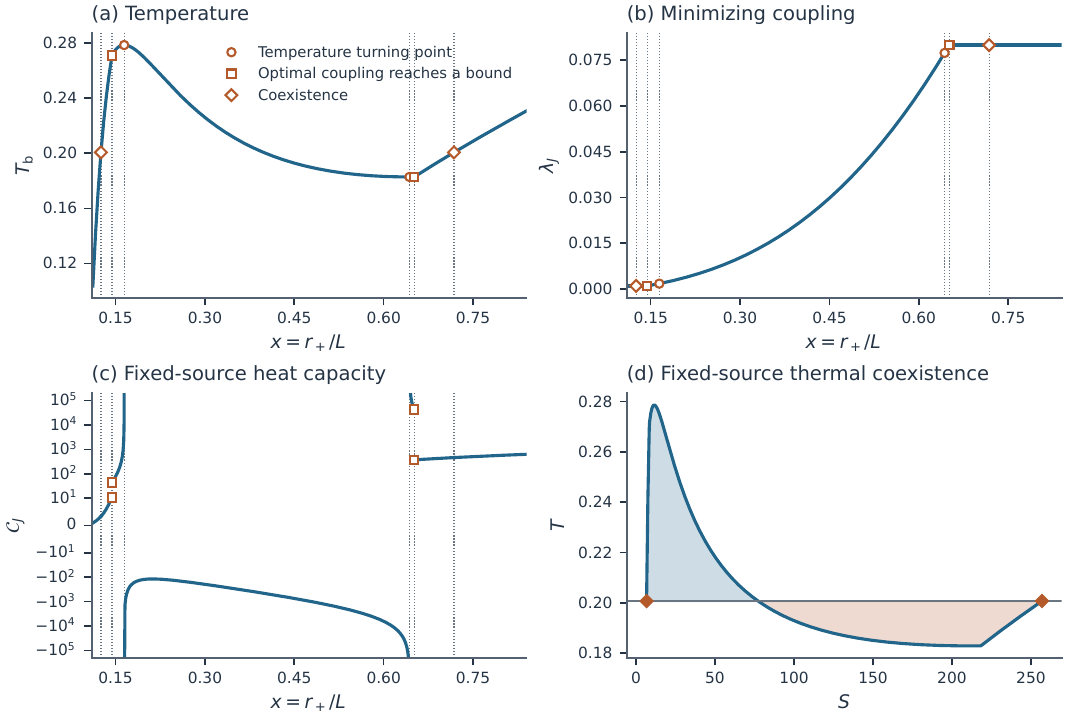}
\caption{Four-dimensional fixed-source diagnostics at \(Q=10,C=2,R=1,J=0.005\) with \(0.001\le\lambda\le0.08\). Panel (a) solves the radial equation \eqref{eq:radial-fixed-source} for the bath temperature at each radius. Panel (b) shows the minimizing coupling given by Eq.~\eqref{eq:lambda} subject to the bounds. Panel (c) evaluates \(\mathcal C_J=\Tb(\dd S/\dd x)/(\dd\Tb/\dd x)\) along this profile using Eq.~\eqref{eq:s4}. Panel (d) compares the state temperature \(T\) with the fixed coexistence bath temperature \(\Tcoex\) in Eq.~\eqref{eq:fixed-source-area} to check equality of the two source potentials. Panels (a)--(c) follow the same stationary profile in radius. Circles mark the temperature folds, squares mark where the minimizing coupling reaches or leaves either bound, as determined by Eq.~\eqref{eq:contact-radius}, and diamonds mark the equal-potential pair. The lower-bound contact occurs at the smaller radius and the higher temperature in this example. Panel (c) uses linear spacing near zero and logarithmic spacing farther away to show both signs of the heat capacity.}
\label{fig:diagnostic}
\end{figure*}

At fixed bath temperature and source, the derivative along the minimizing coupling profile is
\begin{equation}
 \widehat{\Gh}_{J,x}=(T-\Tb)S_x+
        \Gh_{J,\lambda}\,\partial_x\lambda_J .
 \label{eq:radial-profile-derivative}
\end{equation}
The second term vanishes at an interior coupling minimum and on a branch pinned to a fixed bound. Since \(S_x>0\), a radial stationary state on either branch then satisfies
\begin{equation}
 H_J(x,\Tb)=T_4(x,\lambda_J(x,\Tb))-\Tb=0 .
 \label{eq:radial-fixed-source}
\end{equation}
We retain both terms of Eq.~\eqref{eq:radial-profile-derivative} on moving feasibility boundaries. Equation~\eqref{eq:extremal-cap-slope} and the accompanying argument in \ref{app:numerics} exclude minima on the zero-temperature and zero-entropy bounds in the studied ranges. For the regular branches we can determine the stationary temperature uniquely at each radius under a simple condition. At a feasible \(x\) with \(E_\lambda>J\), the unconstrained optimum in Eq.~\eqref{eq:lambda} increases with \(\Tb\). The clipped optimum is therefore nondecreasing in \(\Tb\). Equation~\eqref{eq:t4} decreases with \(\lambda\) at nonnegative state temperature. Hence \(H_J\) strictly decreases with bath temperature. It tends to \(T_4(x,\lambda_-)\) as \(\Tb\to0^+\) and becomes negative at sufficiently large bath temperature. If \(T_4(x,\lambda_-)>0\), Eq.~\eqref{eq:radial-fixed-source} therefore has exactly one positive bath temperature at that radius. This uniqueness is at fixed radius. When the stationary temperature has turning points as a function of radius, a single bath temperature can correspond to several radii, as in Figs.~\ref{fig:four} and \ref{fig:diagnostic}.

A smooth fold is an extremum of this stationary bath temperature as a function of radius. In the interior, \(H_{J,\Tb}<0\) under the same condition. Since \(\widehat{\Gh}_{J,x}=S_x H_J\) there, the fold has zero reduced radial curvature. At a thermal stationary state, \(E_x=TS_x\) gives \(\Gh_{J,xx}=S_x T_x\) and \(\Gh_{J,x\lambda}=S_x T_\lambda\) at fixed bath temperature. Equation~\eqref{eq:reduced} becomes
\begin{equation}
 \widehat{\Gh}_{J,xx}=
 S_x T_x-\frac{(S_xT_\lambda)^2}{\Gh_{J,\lambda\lambda}} .
 \label{eq:fold-coupling-response}
\end{equation}
The second term lowers the radial curvature when the coupling can adjust. Since \(T_\lambda<0\), an interior fold therefore occurs while the same black hole still has positive radial curvature at fixed coupling. At a smooth pinned segment the coupling does not adjust and the radial curvature is \(S_xT_x\).

At both folds of Fig.~\ref{fig:four}, the minimizing coupling is interior. At their numerically determined radii \(x\simeq0.164557\) and \(0.643389\), the fixed-coupling temperature slopes \(RT_x\) are about \(1.357\) and \(0.277\), respectively. The two folds delimit the temperature range with competing minima in this example. The preferred state changes at their equal-potential crossing. Where \(H_{J,\Tb}\ne0\), the stationary temperature slope is \(\Tb'=-H_{J,x}/H_{J,\Tb}\). At fixed \(J,Q,C,R\) the heat capacity is \(\mathcal C_J=\Tb S'/\Tb'\), with both derivatives taken along the stationary profile.

Figure~\ref{fig:diagnostic} separates three events at \(Q=10\). At the larger-radius fold, \(x=0.64338892\), the temperature is stationary and \(\mathcal C_J\) has a pole. The optimal coupling reaches its upper bound farther along the same profile at \(x=0.65132410\). The temperature slope there has positive one-sided limits of about \(0.0031\) and \(0.2775\), so both heat capacities are finite. The coexistence diamonds mark two different radii at one bath temperature. The two heat capacities at coexistence are about \(3.34\) and \(461.62\). At a fold the bath temperature changes only to second order in radius while the entropy still changes to first order. At coexistence, the large-hole branch has both a smaller temperature slope and a larger entropy slope than the small-hole branch. Both effects increase its heat capacity. Between the two folds, entropy increases with radius while temperature decreases. This gives the negative heat capacity of the intervening radial maximum. At a bound contact the coupling stops responding on the pinned side while the temperature retains a finite slope.

Each imposed coupling bound has a unique contact between the interior and pinned portions of the regular stationary curve. At a contact, \(\Xi=J\) and \(\Tb=T_4(x,\lambda)\) at the prescribed endpoint \(\lambda\). Define \(k(x,\lambda)=\log(x/\sqrt{\lambda})-1/2\). The contact radius satisfies
\begin{equation}
 Jx+64\pi^2 C\,xT_4(x,\lambda)k(x,\lambda)=\frac{8\pi C}{R}.
 \label{eq:contact-radius}
\end{equation}
\ref{app:numerics} gives the conditions for uniqueness, which are satisfied for the examples shown here. The upper-bound contact is at a larger radius than the lower-bound contact. Their temperatures depend on the stationary profile. At \(Q=10\), the lower- and upper-bound contact temperatures are about \(0.271132\) and \(0.182826\), respectively. Increasing positive charge moves both contacts outward and lowers their temperatures. These statements concern the stationary curve indexed by radius. Along coexistence, the two states have different radii and the charge at which each one meets a bound also requires equality of their potentials.

\subsection{Charge dependence of the preferred branch}
\label{sec:charge}

\begin{figure*}[t]
\centering
\includegraphics[width=.97\linewidth]{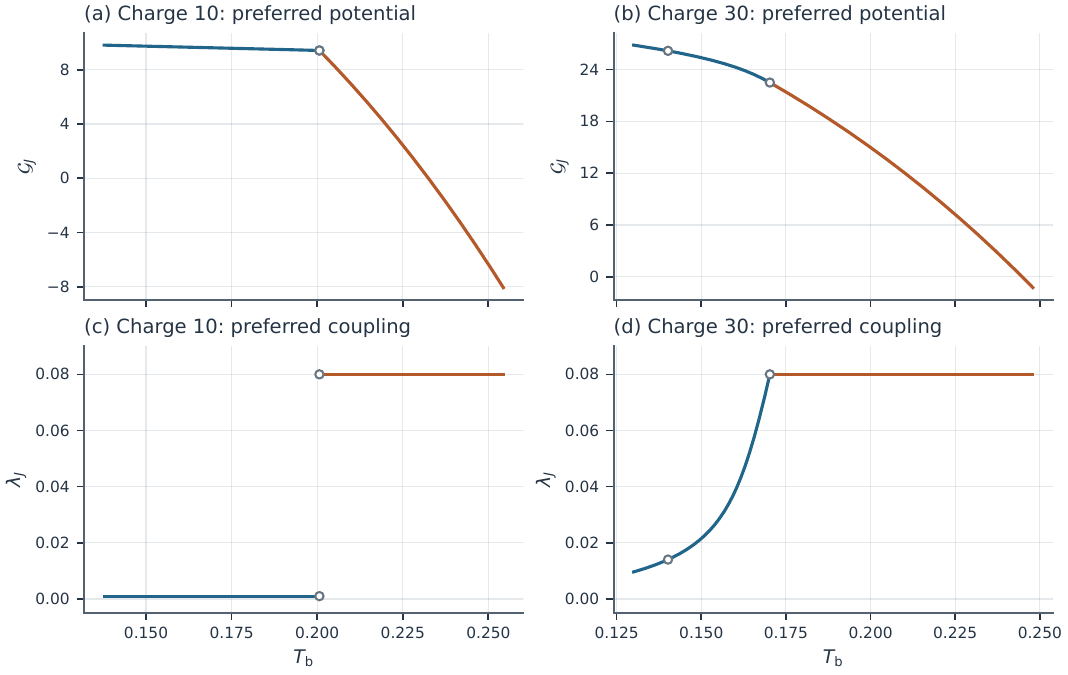}
\caption{Preferred potential and minimizing coupling as bath temperature varies at \(Q=10\) (left) and \(Q=30\) (right). The other controls are given in Sec.~\ref{sec:charge}. The upper panels minimize Eq.~\eqref{eq:sourcepot} and the lower panels show the minimizing coupling from Eq.~\eqref{eq:lambda} subject to the bounds. The left circles mark the two limiting states at the same coexistence temperature. The left panels show a switch between distinct minima at equal potential. The right circles mark an interior solution and the unique upper-bound contact from Eq.~\eqref{eq:contact-radius}, which lies on the preferred branch at this charge. The same minimum continues across the contact. Blue and orange distinguish the portions before and after these events.}
\label{fig:charge}
\end{figure*}

Figure~\ref{fig:charge} compares two charges at the same source \(J=0.005\), with \(C=2,R=1\) and \(0.001\le\lambda\le0.08\). For \(Q=10\), the two preferred branches exchange at \(\Tcoex=0.20067044\). We use subscripts \({\rm s}\) and \({\rm l}\) for the small and large black holes. Their radii are \(0.12561732\) and \(0.71810532\), with couplings at the lower and upper bounds, respectively. The common potential is \(9.42355997\). The temperature derivative of the preferred potential jumps from \(-S_{\rm s}\) to \(-S_{\rm l}\). The intervening stationary state is a radial maximum of higher potential. At fixed \(Q,C,R,J\) and coupling interval, the preferred potential is the infimum of Eq.~\eqref{eq:sourcepot} over a temperature-independent feasible set. It is concave in the bath temperature and continuous wherever it is finite on an open interval. The coupling jump occurs at a kink of this continuous potential.

At \(Q=30\), the global minimum at \(\Tb=0.14020134\) lies at \(x=0.39821472\) and \(\lambda_J=0.01402634\), with positive reduced radial curvature. Following this preferred branch to higher temperature, it reaches \(\lambda_+=0.08\) at \(x\simeq0.663259\) and \(\Tb\simeq0.170282\). The radius, potential, and coupling are continuous there. The first temperature derivative of the potential is also continuous because the entropy approaches the same value from both sides. The coupling changes slope at the contact and then remains at the bound.

\subsection{Coexistence line and critical endpoint}
\label{sec:charge-line}

\begin{figure*}[t]
\centering
\includegraphics[width=.94\linewidth]{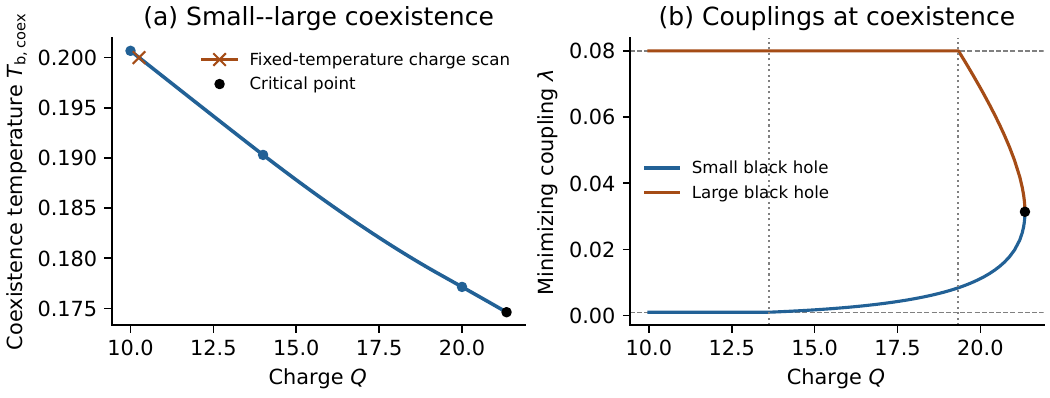}
\caption{Four-dimensional coexistence at \(C=2\), \(R=1\), \(J=0.005\) and \(\lambda\in[0.001,0.08]\). The line equates the two radial minima of Eq.~\eqref{eq:sourcepot} at common controls and has the slope in Eq.~\eqref{eq:charge-line-slope}. Panel (b) uses the coupling minimum in Eq.~\eqref{eq:lambda} subject to the bounds. The black dot in each panel marks the critical endpoint and satisfies \(H_J=\partial_xH_J=\partial_x^2H_J=0\) for Eq.~\eqref{eq:radial-fixed-source}. The blue dots in (a) mark sample coexistence states. The cross marks the equal-potential point on the charge scan at \(\Tb=0.2\), which crosses the coexistence line here. The vertical guides in (b) mark the charges where the two equal-potential branches leave their respective bounds. Their positions are found by solving Eq.~\eqref{eq:contact-radius} numerically together with the coexistence conditions.}
\label{fig:charge-line}
\end{figure*}

We retain the source, size, normalization, and coupling interval of Fig.~\ref{fig:charge} and numerically continue the equal-potential pair in charge. Each point compares minima at a common bath temperature and at that point's fixed charge. Figure~\ref{fig:charge-line} shows the coexistence line and the two minimizing couplings. The small black hole leaves the lower coupling bound at \(Q\simeq13.63292\). The large black hole leaves the upper bound at \(Q\simeq19.33741\). At \(Q=20\), for example, the two black holes coexist at \(\Tb\simeq0.177142\), with minimizing couplings \(0.011137\) and \(0.068763\). Thus the equal-potential comparison can involve two interior couplings. Once a branch leaves its bound, its coupling can adjust so that its thermodynamic response satisfies \(\Xi=J\).

The continuation locates the merger at \(Q_{\rm c}\simeq21.34407\) and \(\Tbc\simeq0.174620\). The critical radius and coupling are \(x_{\rm c}\simeq0.472712\) and \(\lambda_{\rm c}\simeq0.031350\). At this interior critical point, the stationary solution satisfies \(H_J=\partial_x H_J=\partial_x^2 H_J=0\) at fixed bath temperature and charge. The reduced potential has a positive fourth radial derivative there. The radii, entropy, and coupling gaps approach zero along the continued line. The two radial wells and the barrier between them merge into a quartic minimum. The phase distinction therefore disappears at a coupling inside the interval. We checked the equal-potential states against the other radial stationary points and the boundaries of the feasible domain.

The direction of the coexistence line follows from the ordering of the two states. At a common bath temperature, the unconstrained optimum obeys
\begin{equation}
 \partial_x\log\lambda_*=\frac{2}{x}+\frac{1}{4\pi R\Tb x^2}>0.
 \label{eq:coupling-order}
\end{equation}
Consider two regular stationary states at common controls whose couplings are interior or pinned to a fixed endpoint. To compare their entropies, join their radii using \(\lambda_*(x)\) clipped only to the fixed endpoints. This auxiliary curve agrees with the couplings at the two states and is nondecreasing. On its interior segments \(\Tb S_\lambda=E_\lambda-J\) by coupling stationarity. Where it is pinned its coupling derivative vanishes. Since \(S_x>0\), the entropy along this curve increases with radius as long as \(E_\lambda>J\) between the endpoints. For the plotted examples, \(E_\lambda=8\pi C/(Rx)\) exceeds the imposed source throughout the compared radius ranges. Consequently \(S_{\rm l}>S_{\rm s}\) and \(\lambda_{\rm l}\geq\lambda_{\rm s}\). The electric potential has the opposite ordering at positive common charge because \(\Phi=q/(Rx)\).

We use \(\Delta\) for the value on the large black hole minus that on the small black hole, including the potential difference \(\Delta\Gh_J\). Along a regular pair at fixed source, its temperature derivative at fixed charge is \(-\Delta S<0\). Its charge derivative at fixed temperature is \(\Delta\Phi<0\). Either scan therefore has at most one zero for a given pair over a connected interval of regular coexistence candidates. Along the coexistence line, differentiating \(\Delta\Gh_J=0\) at fixed source, volume, and normalization gives
\begin{equation}
 \frac{\dd \Tcoex}{\dd Q}
 =\frac{\Phi_{\rm l}-\Phi_{\rm s}}{S_{\rm l}-S_{\rm s}} .
 \label{eq:charge-line-slope}
\end{equation}
This formula applies to distinct coexisting branches where the entropy difference is nonzero. The two orderings make the slope negative. Raising the charge favors the large hole and lowering the bath temperature restores equality. At \(Q=20\) it gives \(\dd \Tcoex/\dd Q\simeq-0.001860\), in agreement with direct continuation. The fixed-temperature electrical scan in Sec.~\ref{sec:responses} crosses the same line, as indicated in Fig.~\ref{fig:charge-line}(a).

At fixed normalization, varying the source as well as the charge and volume gives
\begin{equation}
 \Delta S\,\dd\Tcoex
   =\Delta\Phi\,\dd Q-\Delta p\,\dd\VV-\Delta\lambda\,\dd J .
 \label{eq:coexistence-differential}
\end{equation}
At fixed \(Q,\VV,C\), the source slope is \(-\Delta\lambda/\Delta S\). It is negative for the two states at distinct couplings here. Increasing the source favors the large hole, so a lower bath temperature restores coexistence. At \(J=0.005\) this derivative is about \(-3.16\times10^{-4}\) at \(Q=10\) and \(-4.00\times10^{-4}\) at \(Q=20\). Direct numerical continuation in \(J\) checks the two values.

The same order of bound contacts and an interior critical point occur at \(J=0.030\) and \(0.100\) with the other controls unchanged. Table~\ref{tab:source-continuation} gives the corresponding contact and critical charges. For all three sources, the interior reduced potential gives square-root closure of the radius, coupling, and entropy gaps as the charge approaches its critical value from below. Appendix~\ref{sec:critical-scaling} derives the leading amplitudes and compares them with the continued solutions.

The same local expansion determines the critical heat-capacity response. At the reference source, tuning the temperature at the critical charge gives a divergence with exponent \(2/3\) in the temperature distance. Evaluating the fixed-charge heat capacity on either coexisting branch instead gives an inverse first power of the charge distance. Appendix~\ref{sec:critical-heat-capacity} gives the amplitudes and compares the two approaches.

\section{Five-dimensional canonical benchmark}
\label{sec:five}

Five-dimensional Gauss--Bonnet thermodynamics exhibits small--large black-hole coexistence. The neutral holographic system has been studied using the Helmholtz swallowtail and the thermal equal-area construction \cite{yangNeutralHolography}. The free-energy branches have also been examined for charged solutions in both four and five bulk dimensions \cite{lyuGaussBonnetThermalBranches}. The charged canonical calculation below provides a benchmark in the normalization of Sec.~\ref{sec:method}.

Each curve fixes \(Q,C,R\) and \(\lambda\). At a given bath temperature only the radius participates in the minimization of \(E_5-\Tb S_5\). Its value on a thermal saddle is \(F=E_5-T_5S_5\). The two locally stable radial branches can coexist at a common bath temperature. Their intervening branch has negative radial curvature. For \(Q=C=10\) and \(\VV=0.01\), the equal-free-energy crossings at \(\lambda=0.005\) and \(0.01\) occur at \(R\Tb=0.45881480\) and \(0.44465777\), respectively.

\begin{figure*}[t]
\centering
\includegraphics[width=.97\linewidth]{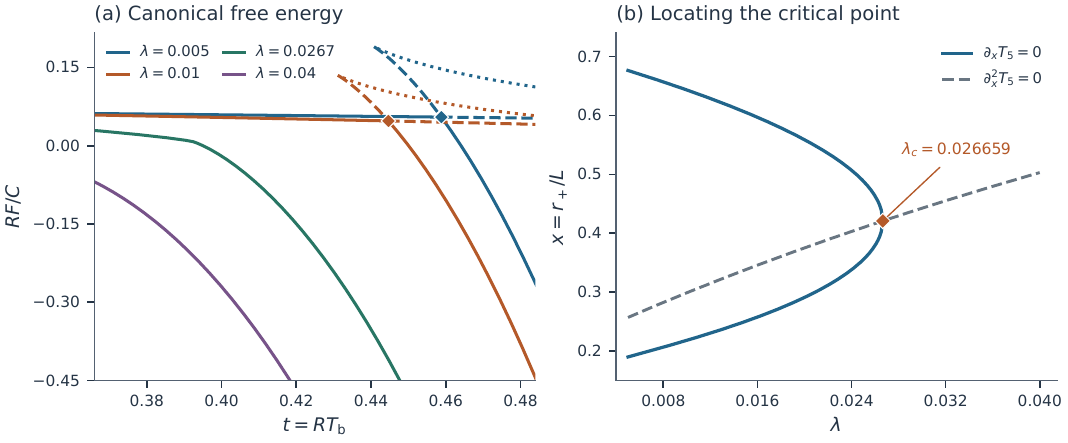}
\caption{Five-dimensional canonical benchmark at \(Q=C=10\) and \(\VV=0.01\). Panel (a) uses \(F=E_5-T_5S_5\) and the state functions in Eqs.~\eqref{eq:e5}--\eqref{eq:t5}. At a thermal saddle the curvature has the sign of \(\partial_xT_5\) because \(\partial_xS_5>0\). In panel (a), solid, dashed, and dotted segments denote preferred radial minima, competing local minima, and radial maxima, respectively. Diamonds identify coexistence at \(\lambda=0.005\) and \(0.01\). Panel (b) uses Eq.~\eqref{eq:t5xx} at fixed \(Q,C,R,\lambda\). The solid curves give the temperature extrema that delimit the range with competing radial minima. The dashed curve marks where the second radial derivative of the temperature vanishes. Their intersection is the merger of the extrema at \(\lambda_c=0.02665899\).}
\label{fig:five}
\end{figure*}

The canonical critical point satisfies \(T_{5,x}=T_{5,xx}=0\), with the coupling held fixed in each derivative. Figure~\ref{fig:five} shows the two extrema of the temperature merging at \(\lambda_c=0.02665899\), \(x_c=0.42092007\), and \(R\Tbc=0.39241425\). At the two illustrated couplings above this charge-specific threshold the temperature profiles are monotone. The extrema bound the temperature window in which two radial minima can compete. A sufficient monotonicity bound for any positive charge follows from the numerator of \(T_{5,x}\) in \ref{app:formulas}. With \(u=x^2\), its charge-independent factor is \(2u^2+(12\lambda-1)u+2\lambda\), whose discriminant is \((36\lambda-1)(4\lambda-1)\). This quadratic is nonnegative for \(1/36\leq\lambda<1/4\), and the remaining charge-dependent term in the derivative numerator is strictly positive. Thus no temperature extrema or canonical swallowtail can occur in that interval at positive charge.

\section{Four-dimensional response diagrams}
\label{sec:responses}

The response curves below show how equilibrium changes as a control is scanned. At imposed pressure \(p_0\) or electric potential \(\Phi_0\), the two candidate states can have different volumes or charges, respectively. The corresponding Legendre potentials were introduced in Sec.~\ref{sec:potentials}. In the fixed-source volume scan, we compare the two candidates at the same volume for each point on the horizontal axis. In the charge scan, the two candidates share the charge at each point. The pressure or electric potential is then a response and can jump when the preferred minimum changes.

\begin{figure*}[t]
\centering
\includegraphics[width=.90\linewidth]{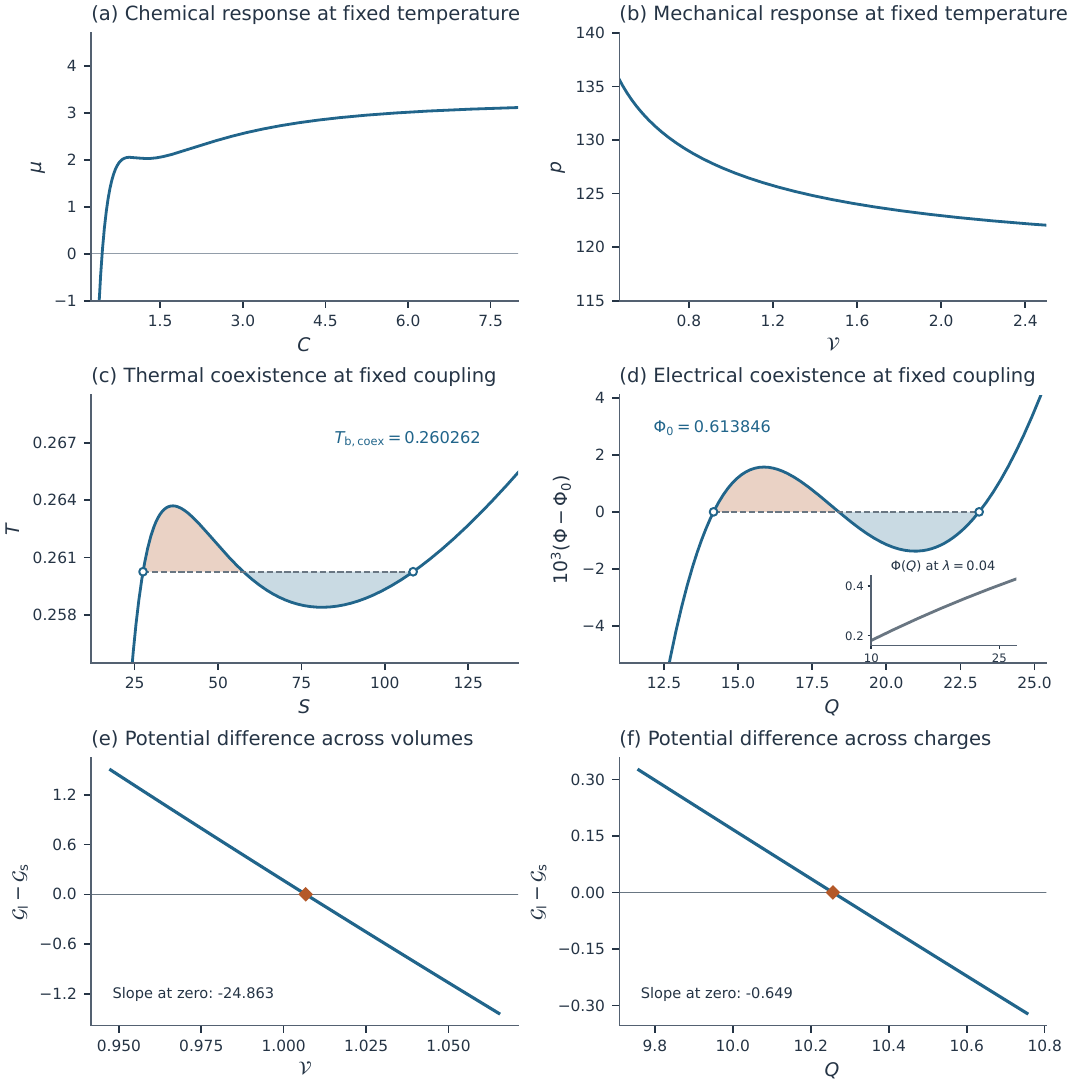}
\caption{Four-dimensional response diagrams in the bare AdS clock. (a) The chemical isotherm at \(\Tb=0.2,\VV=1,Q=10,\lambda=0.005\) uses \(C\mu=E_4-\Tb S_4-\Phi Q\) from Eqs.~\eqref{eq:firstlaw} and \eqref{eq:responses}. The zero and shallow dip occur on a single equilibrium branch. Equations~\eqref{eq:chemical-derivative} and \eqref{eq:chemical-extrema} identify the two extrema of the dip. (b) The pressure at \(\Tb=0.4,C=2,Q=10,\lambda=0.04\) is \(E_4/(2\VV)\) by Eq.~\eqref{eq:responses}. (c) Thermal coexistence is the signed-area zero of Eq.~\eqref{eq:area-t}. (d) Electrical coexistence is the signed-area zero of Eq.~\eqref{eq:area-q}, which gives equal \(F-\Phi_0Q\) at fixed coupling. Both panels use \(\lambda=0.005,C=2,\VV=1\), with \(Q=10\) in (c) and \(\Tb=0.2\) in (d). The inset in (d) shows the monotone \(\lambda=0.04\) isotherm. Panel (e) scans volume and panel (f) scans charge. At each horizontal coordinate the blue curve is the large-minus-small difference of the source potentials in Eq.~\eqref{eq:sourcepot}, evaluated at that same volume or charge for both states. The horizontal line marks zero difference, and the diamond marks equal potentials. The indicated slope is the local derivative at the diamond, given by Eq.~\eqref{eq:dFJ}. Positive and negative differences favor the small and large black holes, respectively. Both panels use \(\Tb=0.2,J=0.005,C=2\) and the bounds of Sec.~\ref{sec:charge}. Panel (e) fixes \(Q=10\), and panel (f) fixes \(R=1\).}
\label{fig:responses}
\end{figure*}

\subsection{Chemical response}

At fixed \(\Tb,Q,\VV,\lambda\), the chemical potential measures how the Helmholtz free energy responds to the boundary normalization \(C\). Increasing \(C\) at fixed total charge reduces \(q=Q/(16\pi C)\). The horizon radius then decreases to maintain the same bath temperature on the isotherm in Fig.~\ref{fig:responses}(a).

At small \(C\) in this plot, the charge contribution to the derivative in Eq.~\eqref{eq:responses} is negative and dominates the other terms. The chemical potential crosses zero at \(C\simeq0.447\). Beyond this crossing, increasing \(C\) raises the equilibrium free energy locally. The shallow dip has a different origin. At fixed radius, the charge energy is proportional to \(1/C\), so its second derivative with respect to \(C\) is positive. Allowing the radius to adjust at fixed temperature reduces the slope of the chemical potential by the term derived in Eq.~\eqref{eq:chemical-slope}. This term dominates between the local maximum at \(C\simeq0.954\) and the local minimum at \(C\simeq1.260\). The entire dip lies on the same smooth equilibrium branch.

At \(\Tb=0.2,C=2,Q=10,R=1,\lambda=0.005\), the chemical potential is \(2.21755873\). Along a path that also varies \(\lambda\) at the same \(\Tb,Q,\VV\), the total derivative of \(F\) with respect to \(C\) also includes \(\Xi\,\dd\lambda/\dd C\).

At fixed nonzero \(Q\) and \(\Tb\), the formal fixed-coupling isotherm extends to \(C\to0\) as the radius increases. \ref{app:numerics} gives the asymptotic relation. Applying the solution in that regime requires an independent assessment of semiclassical validity.

\subsection{Mechanical response}

Boundary pressure and compressibility have been studied through the AdS/CFT correspondence \cite{dolanBoundaryPressure}. For the canonical pressure response here, we vary the boundary volume at fixed \(\Tb,C,Q,\lambda\) and evaluate \(p=E_4/(2\VV)\). Figure~\ref{fig:responses}(b) shows one such isotherm. At \((\lambda,\Tb)=(0.04,0.4)\) and \((0.08,0.3)\), pressure decreases strictly with volume, so Eq.~\eqref{eq:area-p} has no pair of distinct equal-pressure endpoints. \ref{app:numerics} gives the derivative test.

A different mechanical feature can be measured with the volume as the imposed variable. At \(\Tb=0.2\), \(J=0.005\), \(Q=10\), and \(C=2\), the fixed-source minima have equal potentials at the same volume \(\VV=1.00671555\). Their pressures are \(5.33285021\) and \(30.19634464\). The left panel of Fig.~\ref{fig:source-isotherms} marks the two one-sided values. The sign of the jump follows directly from potential equality. For the regular pairs considered above,
\begin{equation}
 \Delta E=\Tb\Delta S+J\Delta\lambda>0
 \label{eq:pressure-order}
\end{equation}
at a crossing with \(J\geq0\). Here differences are large minus small. Since both states have the same volume, \(p=E_4/(2\VV)\) gives \(\Delta p>0\). At every crossing of a connected pair in this range, the volume derivative of its potential difference is \(-\Delta p<0\). Hence the pair can switch only once as volume increases, with pressure jumping upward toward the large branch. The numerical slope at the plotted crossing is \(-24.8634944\). A calculation at imposed \(p_0\) would compare \(F-J\lambda+p_0\VV\) over variable volumes.

\begin{figure}[!htb]
\centering
\includegraphics[width=\linewidth]{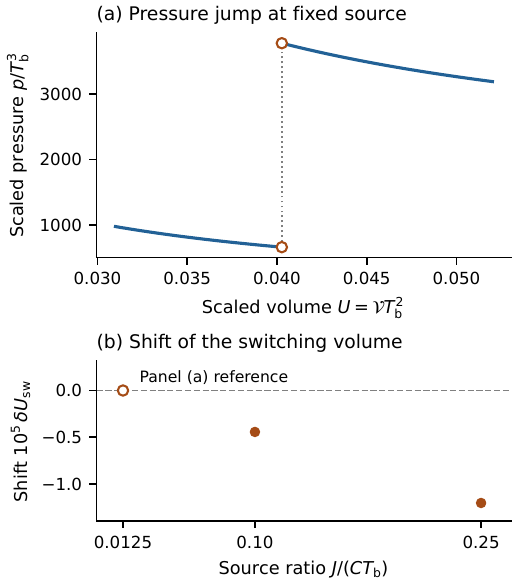}
\caption{Scaled fixed-source mechanical response at \(Q=10\) and \(C=2\). The state functions in Eqs.~\eqref{eq:e4}--\eqref{eq:t4} and the source potential \eqref{eq:sourcepot} give the scaling variables \(\VV\Tb^2\), \(p/\Tb^3\), and \(J/(C\Tb)\), with \(p=E_4/(2\VV)\). (a) At source ratio \(j=J/(C\Tb)=0.0125\), the open circles show the pressures of the two equal-potential states at the same scaled switching volume \(U_{\rm sw}=\VV_{\rm coex}\Tb^2\). The pressure jumps upward as the large black hole becomes preferred. (b) The shift is \(\delta U_{\rm sw}=U_{\rm sw}(j)-U_{\rm sw}(0.0125)\), multiplied by \(10^5\) on the vertical axis. The open point and dashed zero line mark the reference from (a). At larger source ratios the pressure jump occurs at slightly smaller scaled volumes, consistent with Eq.~\eqref{eq:coexistence-differential}.}
\label{fig:mech-scale}
\end{figure}

At fixed \(C,Q\) and coupling bounds, the mechanical isotherms have a simple scaling. The four-dimensional energy is proportional to \(C/R\) and the entropy to \(C\) at fixed radius coordinate and coupling. Dividing the source potential by \(C\Tb\) gives a dimensionless function of \(R\Tb\) and the source ratio \(J/(C\Tb)\). Each isotherm uses a single imposed source \(J\) while the boundary radius varies. Since \(\VV=R^2\), plots of \(p/\Tb^3\) against \(\VV\Tb^2\) coincide at equal \(J/(C\Tb)\). Figure~\ref{fig:mech-scale} shows the scaled pressure jump and the displacement of the switching volume as the source ratio varies. In the displayed range the scaled switching volume changes by only \(1.21\times10^{-5}\). At fixed bath temperature, Eq.~\eqref{eq:coexistence-differential} gives \(\dd\VV_{\rm coex}/\dd J=-\Delta\lambda/\Delta p\), which governs the weak shift. Here the differences are large minus small. The source tilts the relative potentials by the coupling gap, while the sizeable pressure gap makes the potentials cross steeply as volume varies.

\subsection{Thermal response}

At \(Q=10,C=2,\VV=1,\lambda=0.005\), the two regular radial minima have equal Helmholtz potentials at \(\Tcoex=0.26026216\). Their entropies are \(27.51820312\) and \(108.50178849\). Independently solving the signed area equation \eqref{eq:area-t} at fixed coupling gives the same bath temperature. The middle stationary point and the extremal boundary have higher potentials. Figure~\ref{fig:responses}(c) displays the two signed contributions around the common temperature. Their cancellation expresses the balance between the higher energy of the large hole and its greater entropy at the common bath temperature.

For the fixed-source check in Fig.~\ref{fig:diagnostic}(d), we integrate along the stationary branch connecting the two coexisting black holes. The source remains fixed, and the state temperature \(T\) varies along the branch. The identity \(\dd(E-J\lambda)=T\,\dd S\) holds on the interior and fixed-bound pieces because the coupling work cancels. To compare the endpoint potentials at the common bath temperature \(\Tcoex\), we subtract this same temperature throughout the integral. Both endpoints have \(T=\Tcoex\). Splitting at the continuous coupling-bound contacts gives
\begin{equation}
 \Delta\Gh_J(\Tcoex)=\int_{x_{\rm s}}^{x_{\rm l}}
        (T-\Tcoex)S'(x)\,\dd x .
 \label{eq:fixed-source-area}
\end{equation}
At \(\Tcoex=0.20067044\) the endpoint entropies are approximately \(6.72623\) and \(257.02893\). Panel (d) of Fig.~\ref{fig:diagnostic} shows the opposite signed areas. With variable coupling the same balance involves the source-adjusted energy \(E-J\lambda\). Splitting the quadrature at the coupling contacts verifies Eq.~\eqref{eq:fixed-source-area} at the independently found equal-potential temperature.

\subsection{Electrical response}

At imposed electric potential, the phase structure depends on the set of saddles being compared. Small--large coexistence has been studied in the five-dimensional charged Gauss--Bonnet grand canonical ensemble \cite{zhouGaussBonnetGrandCanonical}. A comparison between thermal AdS and a black hole can give a Hawking--Page transition \cite{hawkingPage,wittenThermal}, including in charged Gauss--Bonnet gravity at imposed potential \cite{huGaussBonnetHawkingPage}. Here the electrical test compares the spherical black-hole saddles specified in Sec.~\ref{sec:potentials}.

At \(\Tb=0.2,\VV=1,C=2,\lambda=0.005\), imposing \(\Phi_0=0.61384621\) gives two charged black-hole minima at \(Q=14.17509287\) and \(23.13331537\). They have equal \(F-\Phi_0Q=4.07757309\) in the adopted subtraction. The intervening saddle has potential \(4.08184925\). Minimization over charge is quadratic at fixed radius, so the reduced radial comparison verifies the ordering of these black-hole candidates. An independent solution of Eq.~\eqref{eq:area-q} gives the same imposed potential and endpoint charges. This is the isotherm shown in Fig.~\ref{fig:responses}(d). At the same imposed potential the larger radius carries more charge, as follows from \(\Phi=q/(Rx)\). The electrical work term balances the Helmholtz free-energy difference between these states. At \(\lambda=0.04\) and the same bath temperature, the electric potential increases strictly along the admissible charge scan, as shown in the inset. The same monotonicity holds at \(\lambda=0.10\) for positive temperatures up to \(0.4\).

In the fixed-source charge scan, we vary the imposed charge between comparisons and evaluate \(F-J\lambda\) at the same charge for both candidates in each comparison. At \(\Tb=0.2,\VV=1,C=2,J=0.005\), two black-hole minima exchange preference at the same \(Q=10.25683427\). Their electrical responses are \(\Phi_{\rm s}=0.79165279\) and \(\Phi_{\rm l}=0.14252672\). The slope difference of the two source potentials with respect to \(Q\) agrees with \(\Phi_{\rm l}-\Phi_{\rm s}\). Panels (e) and (f) of Fig.~\ref{fig:responses} show the two differences \(\Gh_{\rm l}-\Gh_{\rm s}\) crossing zero. Their slopes are \(p_{\rm s}-p_{\rm l}\) in the volume scan and \(\Phi_{\rm l}-\Phi_{\rm s}\) in the charge scan. Each zero identifies the imposed volume or charge where the preferred minimum changes.

\begin{figure*}[t]
\centering
\includegraphics[width=.94\linewidth]{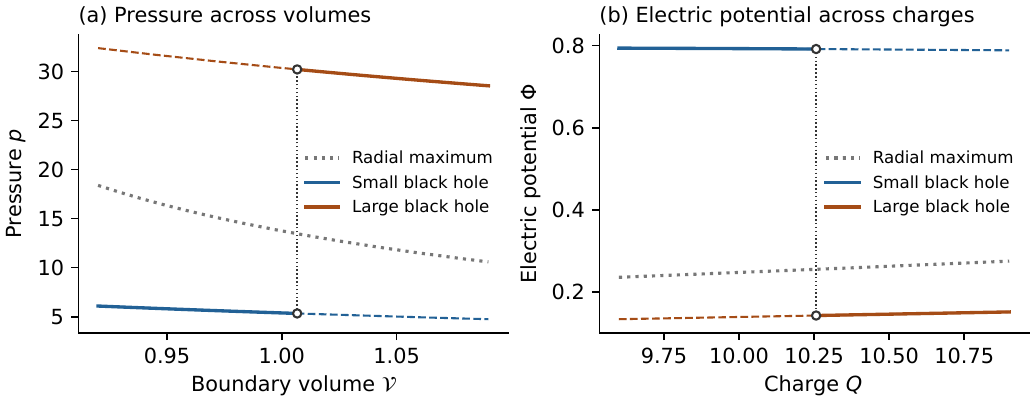}
\caption{Four-dimensional mechanical (left) and electrical (right) isotherms at \(\Tb=0.2\), \(J=0.005\), \(C=2\) and \(\lambda\in[0.001,0.08]\). Each branch solves Eq.~\eqref{eq:radial-fixed-source}. We evaluate \(p=E_4/(2\VV)\) and \(\Phi=q/(Rx)\) using the first-law responses in Eqs.~\eqref{eq:firstlaw} and \eqref{eq:responses}. Curvature distinguishes local extrema, and Eq.~\eqref{eq:sourcepot} ranks the minima. The left panel scans volume at fixed \(Q=10\). The right panel scans charge at fixed \(R=1\). Blue and orange identify the small and large black holes. Solid segments mark their preferred minima, dashed segments their competing local minima, and gray dotted curves the radial maxima. Open circles joined by a vertical guide give the one-sided responses of the two equal-potential states at the same imposed volume or charge.}
\label{fig:source-isotherms}
\end{figure*}

Figure~\ref{fig:source-isotherms} shows how the two potential crossings in Fig.~\ref{fig:responses}(e,f) appear in the equations of state. At each crossing, the vertical guide joins the two responses at the same imposed volume or charge. Changing from the small to the large black hole raises the pressure in the volume scan and lowers the electric potential in the charge scan.

\section{Conclusion}

At positive bath temperature the four-dimensional logarithmic entropy allows exact minimization over the coupling at each radius. Allowing an interior coupling to adjust lowers the radial curvature and accounts for the temperature folds in the reference example. The preferred potential distinguishes two outcomes. Two distinct minima can exchange preference at equal potential, producing jumps in the equilibrium responses. In the higher-charge example a single preferred minimum reaches a coupling bound continuously, and the first temperature derivative of the preferred potential is continuous there.

At the three sources studied, numerical continuation shows that the coexisting minima leave their respective coupling bounds in succession and merge at an interior critical point. Expanding the reduced potential around the merger gives square-root closure of the radius, coupling, and entropy gaps. At the reference source this expansion also gives distinct heat-capacity powers for a temperature sweep at the critical charge and an approach along coexistence. The ordering of the two states accounts for the negative charge--temperature coexistence slope, the upward pressure jump in the volume scan, and the downward electric-potential jump in the charge scan.

In five dimensions, the fixed-coupling benchmark ends at the merger of the two radial temperature extrema. The fixed-source calculation in \ref{app:five-source} instead allows the coupling to vary. It gives coexistence between the coupling endpoints and compares the required source when the common bath temperature is specified in bare-length time or in proper cylinder time.

\begin{acknowledgements}
Limin Zeng would like to express his deepest and foremost thanks to his partner Lu Lu.
\end{acknowledgements}

\appendix
\renewcommand{\thesubsection}{\Alph{section}.\arabic{subsection}}
\renewcommand{\theequation}{\Alph{section}.\arabic{equation}}
\renewcommand{\theHequation}{appendix.\Alph{section}.\arabic{equation}}
\section{Geometric and derivative checks}
\label{app:formulas}

\subsection{Horizon equation and dimensionful variation}

For the spherical metric function \(f(r)=1-r^2\psi(r)\), the Einstein branch satisfies
\begin{equation}
 \psi+\alpha\psi^2=-L^{-2}+\frac{m}{r^{d-1}}
                         -\frac{a_Q}{r^{2d-4}} .
 \label{eq:metric}
\end{equation}
At infinity the mass and charge terms decay. Substituting \(\psi=-1/L_{\rm eff}^2\) into the vacuum equation and taking the root continuous at \(\alpha=0\) gives Eq.~\eqref{eq:clock-conversion}. The horizon condition \(f(r_+)=0\) fixes the mass parameter to
\begin{equation}
 m=r_+^{d-3}+\frac{r_+^{d-1}}{L^2}
     +\alpha r_+^{d-5}+\frac{a_Q}{r_+^{d-3}} .
 \label{eq:horizon-mass}
\end{equation}
In four dimensions \((m,a_Q)=(2GM,GQ_b^2)\). In five dimensions \((m,a_Q)=(16\pi GM/3,GQ_b^2/3)\) in the stated volume normalization. The energy follows from \(E=ML/R\). Differentiating Eq.~\eqref{eq:metric} at the horizon gives \(T=(L/R)f'(r_+)/(4\pi)\). The five-dimensional entropy is the sum of the Wald area and curvature terms \cite{wald,jacobsonMyers,cai}.

In four dimensions, keep \(L,G,Q_b,\alpha\) fixed and use the surface-gravity temperature \(T_H=f'(r_+)/(4\pi)\). The radial first law \(M_{r_+}=T_H S_{r_+}\) then gives the same entropy derivative in boundary variables,
\begin{equation*}
 \partial_x S_4=32\pi^2 C\left(x+\frac{2\lambda}{x}\right).
\end{equation*}
The \(1/x\) term integrates to a logarithm \cite{fernandesCharged4d,wei}. The radial law leaves an integration term independent of \(x\). Fixing the reference length to \(\sqrt{\alpha}\) gives Eq.~\eqref{eq:s4}, which is the entropy used in Eq.~\eqref{eq:sourcepot}.

We use tildes here for the variables of the dimensionful coupling description. This coordinate uses the actual bare AdS length, so \(\aold=L^{d-2}\lambda=16\pi GC\lambda\). The chain rule in Eq.~\eqref{eq:firstlaw} gives
\begin{align}
 \dd E&=T\,\dd S-p\,\dd\VV+\Phi\,\dd Q
       +\widetilde\mu\,\dd C \nonumber\\
 &\quad+\Aold\,\dd\aold-\lambda\Xi\,\frac{\dd G}{G},\label{eq:dimensionful}\\
 \widetilde\mu&=\mu-\frac{\lambda\Xi}{C},\qquad
 \Aold=\frac{\Xi}{L^{d-2}}.\nonumber
\end{align}
The response \(\Aold\) is read from the state first law in this coordinate, and \(\widetilde\mu\) is the chemical potential for variations of \(C\) at fixed \(S,Q,\VV,G,\aold\). The final term vanishes on fixed-\(G\) variations. In Eq.~\eqref{eq:firstlaw} the entropy, charge, normalization, boundary radius, and dimensionless coupling are independent.

The linear control in Eq.~\eqref{eq:sourcepot} can also be written using a dimensionful reference coordinate. Choose an external length \(L_*\) that is fixed across all candidates. Then \(\alpha_{\rm ref}=L_*^{d-2}\lambda\) and \(J_{\rm ref}=J/L_*^{d-2}\), so \(J_{\rm ref}\alpha_{\rm ref}=J\lambda\). This change of units preserves the control term. The coordinate \(\aold\) in Eq.~\eqref{eq:dimensionful} uses the actual \(L\) of each bulk representative.

\subsection{Temperature derivatives}

At fixed coupling the denominator of \(T_{4,x}\) is \(4\pi R D_4^2\), where \(D_4=x(x^2+2\lambda)\), and the numerator is
\begin{equation}
 P_4=3x^6+(18\lambda-1)x^4+(5\lambda+3q^2)x^2
                +2\lambda(\lambda+q^2).
\end{equation}
The second derivative of \(T_4\) has numerator \(D_4 P'_4-2D'_4P_4\) after clearing positive factors. On a fixed-source radial trace, the relevant derivative is the total derivative of \(T(x,\lambda_J(x,\Tb))\) subject to the temperature equation. The radial and coupling responses both enter this derivative. In the fixed-source example we check the curvature of the reduced potential as well as the temperature derivative.

For the five-dimensional curve let \(u=x^2\) and define
\begin{align}
 P_5(u)&=3u^2[2u^2+(12\lambda-1)u+2\lambda]+q^2(5u+6\lambda),\\
 H_5(u)&=u(u+2\lambda)P'_5(u)-4(u+\lambda)P_5(u).
\end{align}
With fixed \(Q,C,R,\lambda\), the first and second derivatives of \(T_5\) are
\begin{align}
 T_{5,x}&=\frac{P_5(u)}{6\pi R x^4(x^2+2\lambda)^2},\\
 T_{5,xx}&=\frac{H_5(u)}{3\pi R x^5(x^2+2\lambda)^3}.
 \label{eq:t5xx}
\end{align}
For example, at \(Q=14,C=10,R=1,\lambda=0.02,x=0.45\), both direct differentiation of Eq.~\eqref{eq:t5} and Eq.~\eqref{eq:t5xx} give \(T_{5,xx}=0.523909788586\).

\section{Admissibility and reproduction of the response checks}
\label{app:numerics}

\subsection{Feasible domain, minima, and coupling bounds}

In the four-dimensional fixed-source example, the interval is contained in \(0<\lambda<1/4\), where the Einstein-branch AdS vacuum has a real curvature scale. We use positive-temperature, positive-entropy outer horizons. For each fixed radius we allow the closure of these constraints when testing the boundary values of the potential. The nonnegative-temperature condition is \(\lambda\leq3x^4+x^2-q^2\). For the adopted entropy reference, \(S_4\geq0\) is equivalent to \(x^2\geq\beta_S\lambda\), with \(\beta_S=2W_0(1/2)\) and \(W_0\) the principal Lambert function. Thus the allowed upper bound at fixed \(x\) is
\begin{equation}
 \lambda_{\rm cap}(x)=\min\{\lambda_+,\,3x^4+x^2-q^2,\,x^2/\beta_S\}.
 \label{eq:feasible-coupling-cap}
\end{equation}
Radii with \(\lambda_{\rm cap}<\lambda_-\) are excluded. The feasible set is determined by the geometry and the entropy prescription at fixed charge and normalization. If \(E_\lambda>J\), Eq.~\eqref{eq:lambda} gives \(\lambda_*<x^2/e\). The zero-entropy bound \(x^2/\beta_S\) is higher than this unconstrained optimum, so it cannot be active at the minimum when the feasible interval has nonzero width. At the plotted settings, \(E_\lambda>J\) throughout the radial search. A sufficient condition for \(x\leq3\) is \(J<8\pi C/(3R)\).

On an active zero-temperature bound \(\lambda_{\rm ext}(x)=3x^4+x^2-q^2\), the state temperature vanishes, \(\lambda_{\rm ext}'(x)>0\), and \(\Gh_{J,\lambda}\leq0\). Equation~\eqref{eq:radial-profile-derivative} then gives
\begin{equation}
 \widehat{\Gh}_{J,x}
 =-\Tb S_x+\Gh_{J,\lambda}\lambda_{\rm ext}'(x)<0 .
 \label{eq:extremal-cap-slope}
\end{equation}
The potential also decreases as this boundary meets a fixed upper bound at zero temperature, since the pinned-side derivative is \(-\Tb S_x<0\) there. The leftmost feasible radius has \(\lambda=\lambda_-\). At the zero-entropy radius for this coupling, the numerator of \(T_4\) is
\begin{equation*}
 \lambda_-\bigl(3\beta_S^2\lambda_-+\beta_S-1\bigr)-q^2 .
\end{equation*}
It is negative if \(\lambda_-<(1-\beta_S)/(3\beta_S^2)\simeq0.19974\), as in the examples. Thus the left edge is extremal. A trial with \(\lambda=\lambda_-\) and slightly larger \(x\) lowers the potential there. The global comparison in the plotted cases therefore reduces to regular positive-temperature stationary states, including any at fixed-bound contacts.

To prove uniqueness in Eq.~\eqref{eq:contact-radius}, let \(x_0\) be the larger of the zero-temperature radius at the given endpoint and \(\sqrt{e\lambda}\). Assume \(0\leq J<8\pi C/(R x_0)\). At positive temperature below \(x_0\), the logarithmic factor \(k\) is nonpositive, so Eq.~\eqref{eq:contact-radius} cannot hold. At \(x_0\) the product \(xT_4 k\) vanishes and the left side is below the right side. For \(x>x_0\), \(k\) strictly increases and \(xT_4\) strictly increases. In terms of \(u=x^2\), the derivative of \(4\pi R xT_4\) with respect to \(u\) has numerator \(3u^2+12\lambda u+3\lambda+q^2\) over \((u+2\lambda)^2\). The contact equation therefore has one root, since its left side increases without bound. That root has positive temperature and entropy.

At the root, increasing either the charge or the endpoint coupling decreases \(xT_4 k\) at fixed \(x\). For charge this follows from \(T_{4,Q}<0\) at positive \(Q\). For the coupling, both \(T_4\) and \(k\) decrease at fixed \(x\). The root therefore moves outward in both cases. For a fixed coupling endpoint, the contact temperature from Eq.~\eqref{eq:contact-radius} is proportional to \((8\pi C/(Rx)-J)/k\), which decreases with \(x\) at the contact. Thus increasing charge lowers the contact temperature. Solving the scalar equation for \(Q=10\) gives \(x\simeq0.144225\) at the lower bound and \(0.651324\) at the upper bound, as in Fig.~\ref{fig:diagnostic}.

For the numerical checks, we split the reduced radial profile at the contacts and compare stationary and boundary values using Eq.~\eqref{eq:radial-profile-derivative}. The regular search brackets sign changes of the radial derivative. The folds are located separately from \(H_J=\partial_xH_J=0\) on smooth branches. The critical point is obtained by also imposing \(\partial_x^2H_J=0\).

The radial search extends to \(x=3\). For \(x\geq3\), \(\lambda\leq0.10\), and \(q^2\leq1\), Eq.~\eqref{eq:t4} gives \(RT_4\geq2x/(5\pi)\). The bath temperatures in the four-dimensional fixed-source searches obey \(R\Tb<6/(5\pi)\). Consequently the potential at every allowed fixed coupling increases beyond the cutoff, and all couplings are feasible there.

At a fixed coupling a term in the entropy that is independent of radius cancels from same-temperature potential differences. Across couplings such a term contributes a coupling-dependent offset to the potential. We therefore keep the reference in Eq.~\eqref{eq:s4} fixed throughout the comparison.

When the two coexisting branches continue to lie at the opposite coupling bounds, their potential difference obeys
\begin{equation}
 \Delta S\,\dd\Tb=(\Xi_{\rm l}-J)\dd\lambda_+
                      -(\Xi_{\rm s}-J)\dd\lambda_- ,
 \label{eq:boundaries}
\end{equation}
at fixed \(Q,C,R,J\), where \(\Delta S=S_{\rm l}-S_{\rm s}\). The formula follows by differentiating each stationary potential and imposing equality. At the reference coexistence, the derivatives with respect to the upper and lower bounds are \(-0.15762439\) and \(-0.7080088\), respectively. Central differences agree with these values. Raising the upper bound lets the large branch lower its potential further. Raising the lower bound raises the potential of the small branch. Each change favors the large black hole at the original coexistence temperature. With the lower bound fixed at \(0.001\), moving the upper bound to \(0.06\) or \(0.10\) gives the respective equal-potential temperatures \(0.20474097\) and \(0.19814127\).

\subsection{Source continuation and local critical scaling}
\label{sec:critical-scaling}

Table~\ref{tab:source-continuation} records the charges at which the two coexisting states leave their respective coupling bounds for each of the three source values. Equation~\eqref{eq:coexistence-differential} gives the temperature shift with source at fixed charge and volume for distinct coexisting states. For each sampled charge, the radial search compares the continued minima with the other stationary points and all boundaries of the feasible domain. The contact and merger charges in the table occur in the same order for these three sources. For the sampled sources, both branches can thus adjust their couplings in the interior before the two minima merge.

\begin{table}[t]
\caption{Source dependence at the fixed size, normalization, and coupling interval of Fig.~\ref{fig:charge-line}. Here \(Q_{\rm s}\) and \(Q_{\rm l}\) mark where the small and large black holes leave their respective coupling bounds.}
\label{tab:source-continuation}
\centering
\begin{tabular}{@{}rrrrr@{}}
\toprule
\(J\) & \(Q_{\rm s}\) & \(Q_{\rm l}\) & \(Q_{\rm c}\) & \(\Tbc\) \\
\midrule
0.005 & 13.63292 & 19.33741 & 21.34407 & 0.17462035 \\
0.030 & 13.63287 & 19.33947 & 21.34502 & 0.17460877 \\
0.100 & 13.63272 & 19.34522 & 21.34768 & 0.17457632 \\
\bottomrule
\end{tabular}
\end{table}

Write \(\widehat{\Gh}\) for the smooth interior reduced potential at fixed \(J\). Its first three radial derivatives vanish at the critical point. At the numerically located critical point for each source in Table~\ref{tab:source-continuation}, the mixed derivative \(\widehat{\Gh}_{x\Tb}\) is nonzero. The limiting slope \(\tau_c=-\widehat{\Gh}_{xQ}/\widehat{\Gh}_{x\Tb}\) follows from Eq.~\eqref{eq:charge-line-slope} as the two radii merge. Along that tangent the linear variation of the radial force in \(Q\) and \(\Tb\) cancels. Define \(A_c=\widehat{\Gh}_{xxQ}+\tau_c\widehat{\Gh}_{xx\Tb}\) and \(B_c=\widehat{\Gh}_{xxxx}\) at the critical point. Both are positive for the tabulated sources. With \(\epsilon=Q_{\rm c}-Q\) and \(y=x-x_{\rm c}\), the leading quadratic and quartic terms are
\begin{equation*}
 -\tfrac12 A_c\epsilon y^2 + \tfrac1{24} B_c y^4 .
\end{equation*}
A linear term of the same order would tilt this double well and split the two depths. Equality of the depths on the coexistence line fixes that leading tilt to zero. The separation of the two minima is therefore
\begin{equation}
 \Delta x=2\sqrt{6A_c/B_c}\,\sqrt{Q_{\rm c}-Q}
          +o\!\left(\sqrt{Q_{\rm c}-Q}\right).
 \label{eq:critical-square-root}
\end{equation}
The minimizing coupling and the entropy along the reduced potential are smooth functions of radius at the critical point. Their first radial derivatives are nonzero, so their jumps have the same power. At \(J=0.005\) the leading amplitudes of \(\Delta x\), \(\Delta\lambda\), and \(\Delta S\) are \(0.243956\), \(0.047956\), and \(122.483\). At a charge distance of \(0.0005\) the relative discrepancy between the three numerical gaps and their leading expressions is at most \(1.4\times10^{-5}\) for the sources in the table. Solving the zero-gradient and null-curvature conditions in the full two-variable potential, including the vanishing third derivative along its null direction, gives the same critical points.

\subsection{Critical heat capacity}
\label{sec:critical-heat-capacity}

\begin{figure*}[t]
\centering
\includegraphics[width=.90\linewidth]{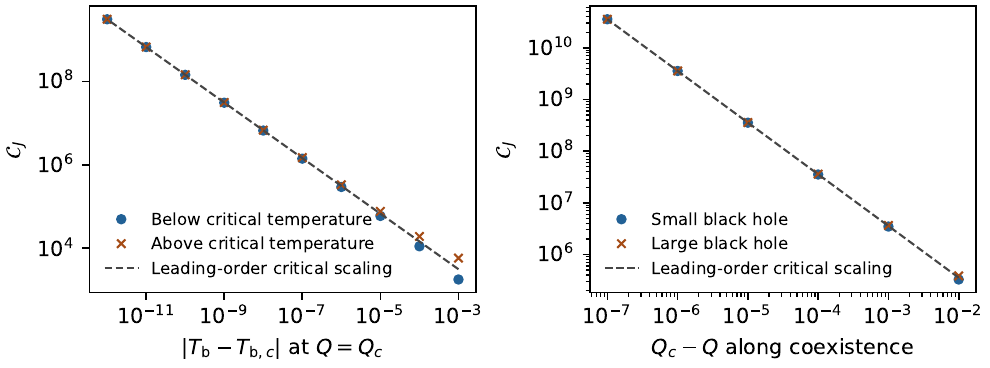}
\caption{Critical heat capacity at the reference source \(J=0.005\). The points use the full radius--coupling Hessian of Eq.~\eqref{eq:sourcepot}; the equivalent reduced derivative is Eq.~\eqref{eq:reduced-heat-capacity}. On the left, the labels distinguish bath temperatures below and above the critical temperature, at fixed critical charge. On the right, the labels distinguish the small and large black holes along the coexistence line as their common charge approaches its critical value. Dashed lines show the leading-order critical scaling from Eqs.~\eqref{eq:heat-isocharge} and \eqref{eq:heat-coexistence}.}
\label{fig:critical-heat-capacity}
\end{figure*}

On a regular radial minimum with \(\widehat{\Gh}_x=0\) and \(\widehat{\Gh}_{xx}>0\), the equilibrium entropy is \(-\widehat{\Gh}_{\Tb}\). The fixed-charge heat capacity at fixed source, size, and normalization follows by differentiating along the radial minimum,
\begin{equation}
 \mathcal C_J=\Tb\left(\frac{\widehat{\Gh}_{x\Tb}^2}{\widehat{\Gh}_{xx}}-\widehat{\Gh}_{\Tb\Tb}\right).
 \label{eq:reduced-heat-capacity}
\end{equation}
Let \(D_c=\widehat{\Gh}_{x\Tb}|_{\rm c}\), and retain \(A_c,B_c\) from Appendix~\ref{sec:critical-scaling}. The reference critical point has \(D_c\ne0\). At fixed critical charge, the radial equation balances its linear temperature variation against its cubic radial term. The displacement is therefore proportional to the cube root of the temperature distance. The radial curvature vanishes quadratically in this displacement, yielding
\begin{equation}
 \mathcal C_J\sim\frac{2\Tbc D_c^2}{B_c(6|D_c|/B_c)^{2/3}}
 |\Tb-\Tbc|^{-2/3}.
 \label{eq:heat-isocharge}
\end{equation}
Along the coexistence line, Eq.~\eqref{eq:critical-square-root} gives \(\widehat{\Gh}_{xx}\sim 2A_c(Q_{\rm c}-Q)\) on each minimum. Equation~\eqref{eq:reduced-heat-capacity} then gives
\begin{equation}
 \mathcal C_J\sim\frac{\Tbc D_c^2}{2A_c}(Q_{\rm c}-Q)^{-1}.
 \label{eq:heat-coexistence}
\end{equation}
Here each heat capacity is a temperature derivative at the charge of the corresponding coexistence point. The two equations describe different paths to the same critical point. Both divergences reflect the vanishing radial curvature at the same critical configuration after the coupling has adjusted. The two sides of the temperature sweep have the same leading amplitude. The two coexisting branches also share a leading amplitude, which explains the overlap in each panel close to criticality. A temperature sweep tilts the quartic potential, whereas approaching along coexistence makes the two equal-depth wells progressively shallower. At \(J=0.005\) the leading amplitudes are \(31.1302\) and \(3561.21\), respectively, in the normalization used here. Figure~\ref{fig:critical-heat-capacity} compares both laws with heat capacities computed independently from the full radius--coupling Hessian at the numerical saddles.

\subsection{Isotherms and sign tests}

The four-dimensional isotherm equation at fixed coupling is
\begin{equation}
 q^2=U(x)=3x^4+x^2-\lambda-4\pi R\Tb x(x^2+2\lambda).
 \label{eq:isotherm-charge}
\end{equation}
The chemical isotherm uses \(C=Q/(16\pi\sqrt U)\) at \(Q>0\). At large \(x\), \(U\sim3x^4\), so the formal \(C\) tends to zero. At the parameters of Fig.~\ref{fig:responses}(a), \(U(0)<0\), \(U'(0)<0\), and \(U''(x)>0\), so \(U(x)=q^2>0\) has exactly one positive root. The temperature rises with \(x\) at that root, giving a radial minimum. This establishes that the shallow dip occurs on a single equilibrium branch.

Write \(\mathcal F_4(x,C)=E_4-\Tb S_4\) at fixed \(\Tb,Q,\VV,\lambda\). Differentiating its stationary-radius condition with respect to \(C\) gives
\begin{equation}
 \frac{\dd\mu}{\dd C}
   =\mathcal F_{4,CC}
    -\frac{\mathcal F_{4,xC}^2}{\mathcal F_{4,xx}} .
 \label{eq:chemical-slope}
\end{equation}
Here \(E_{\rm el}\) denotes the electric term in Eq.~\eqref{eq:e4}. At fixed radius all the other terms are linear in \(C\), so
\begin{equation*}
 \mathcal F_{4,CC}=\frac{2E_{\rm el}}{C^2},\qquad
 E_{\rm el}=\frac{Q^2}{32\pi R C x}.
\end{equation*}
The second term in Eq.~\eqref{eq:chemical-slope} reduces the slope as the radius adjusts. The number of extrema can also be obtained without tracing \(\mu(C)\) numerically. Substitute \(q^2=U(x)\) into Eq.~\eqref{eq:responses} and differentiate along the isotherm:
\begin{align}
 \frac{\dd\mu}{\dd x}&=-\frac{16\pi}{R x^2}P_\mu(x),\label{eq:chemical-derivative}\\
 P_\mu(x)&=3x^4-2\pi R\Tb x^3+4\pi R\Tb\lambda x+\lambda .
 \label{eq:chemical-extrema}
\end{align}
The nonzero coefficients have two sign changes, so Descartes' rule allows at most two positive roots. At the controls of Fig.~\ref{fig:responses}(a), the bounds \(3.14<\pi<22/7\) fix the signs at \(x=0.25,0.30,0.40\) as positive, negative, and positive. This interval lies on the admissible isotherm with \(U>0\) and \(U'>0\). Both positive roots therefore occur on the branch. Since \(C(x)\) strictly decreases there, they are precisely the local maximum and minimum of the chemical response. Numerical evaluation of the roots gives their positions quoted in the main text.

For the mechanical isotherm, define \(t(x)=RT_4(x)\) at fixed \(q,\lambda\). At positive imposed \(\Tb\), solve \(R=t/\Tb\) and then compute \(p=E_4/(2R^2)\). Write \(u=x^2\) and \(h=q^2\). Up to a positive constant independent of \(x\), the pressure is \(P_p/N_p^3\), where
\begin{align}
 N_p&=3u^2+u-\lambda-h,\\
 P_p&=(u^2+u+\lambda+h)\,u(u+2\lambda)^3 .
\end{align}
For the mechanical examples at \(C=2,Q=10\), \(h>0.0098\). The coefficients of \(P_p'N_p-3P_pN_p'\) are nonincreasing functions of \(h\) for \(h>0\). At \(h=0.0098\), the lower powers of \(u\) have negative coefficients. The three highest powers, after dividing by \(u^4\), give a quadratic with negative leading coefficient and negative discriminant, for both \(\lambda=0.04\) and \(0.08\). The derivative of pressure is strictly negative on \(N_p>0\). In both cases \(P_4\) is positive on \(u>0\), so \(R\) and \(\VV\) increase with radius. This proves the pressure monotonicity used in Sec.~\ref{sec:responses}.

Consider the electrical isotherm at \(R=1\). Writing \(\Phi^2=U/x^2\) using Eq.~\eqref{eq:isotherm-charge}, we obtain the numerator of its radial derivative,
\begin{equation}
 P_\Phi(x)=6x^4-4\pi\Tb x(x^2-2\lambda)+2\lambda .
\end{equation}
On \(x^2\leq2\lambda\), this expression is positive for \(\Tb>0,\lambda>0\). On \(x^2>2\lambda\), replace \(\pi\) by \(22/7\) and \(\Tb\) by the maximum temperature under consideration to obtain a lower bound. At \((\lambda,T_{{\rm b},\max})=(0.04,0.2)\) and \((0.10,0.4)\), the resulting rational-coefficient polynomials have no positive roots by exact root counting, and their constants are positive. Thus \(\Phi\) increases on the admissible positive-charge isotherm. Since \(Q=16\pi C x\Phi\) at \(R=1\), charge increases there as well. This rules out distinct equal-potential endpoints in the monotone cases discussed in Sec.~\ref{sec:responses}.

\subsection{Endpoint and integral reproduction}

For the canonical crossings we solve \(T(x_{\rm s})=T(x_{\rm l})=\Tb\) and equality of \(E-\Tb S\). We independently solve the same temperature equations with the signed area in Eq.~\eqref{eq:area-t}. At a root, the radial Hessian of \(E-\Tb S\) is \(S_xT_x\), with \(S_x>0\). The two stable candidates are the outer roots of the temperature equation. The intervening candidate and extremal boundary have higher potentials at each reported crossing. The electrical check solves the same-potential endpoint equations and independently solves Eq.~\eqref{eq:area-q}. At imposed \(\Phi_0\), the minimizing charge at a fixed radius is \(Q=16\pi C R x\Phi_0\). Substituting it into \(E-\Tb S-\Phi_0Q\) yields a radial potential that grows both at small and large radius for the example in Sec.~\ref{sec:responses}. Its three positive stationary radii are approximately \(0.22970290\), \(0.29811358\), and \(0.37486807\). The outer radii give equal potentials and positive reduced curvatures, while the middle point has negative curvature.

The endpoint and signed-area checks are implemented separately in \texttt{numerics/verify.py} in the reproducibility archive~\cite{zengReproducibility}. The endpoint and integral residuals are saved in \texttt{results.json}. The plotting scripts read the numerical tables and draw each curve in radial order on its own continuous branch. The archive's README maps the manuscript figures to their scripts and gives the command for rerunning the analytic checks and regenerating the numerical inputs.

\subsection{Comparison with the published diagrams}

Table~\ref{tab:published-diagrams} relates the results to the questions addressed by the figure sets of Ref.~\cite{epjc}. The five-dimensional main-text comparison fixes the coupling on each curve. \ref{app:five-source} allows the coupling to vary at fixed source.

\begin{table}[t]
\caption{Correspondence with the figure sets of Ref.~\cite{epjc}.}
\label{tab:published-diagrams}
\begin{tabular}{@{}p{.23\linewidth}p{.71\linewidth}@{}}
\toprule
Published figures & Calculation in this work \\
\midrule
1--10 & Four-dimensional fixed-source radial branches, heat capacity, and charge continuation \\
11--12 & Five-dimensional fixed-coupling benchmark and fixed-source endpoint comparison \\
13--14 & Chemical isotherm at fixed coupling \\
15--17 & Fixed-coupling pressure and scaled fixed-source isotherms \\
18 & Signed-area test at fixed coupling and at fixed source \\
19--21 & Imposed-potential coexistence at fixed coupling and imposed-charge switch at fixed source \\
\bottomrule
\end{tabular}
\end{table}

\section{Five-dimensional endpoint comparison at fixed source}
\label{app:five-source}

We consider the coupling interval \([0,\lambda_U]\), with \(0<\lambda_U<1/4\). Its endpoints are the Einstein value \(\lambda=0\) and the prescribed upper bound \(\lambda=\lambda_U\). We start with the bare-length clock and background subtraction. At fixed \(C,R,Q\), we minimize over radius and coupling at a common bath temperature and source \(J\). In units of \(C/R\), the reduced source is \(z=RJ/C\). With \(t=R\Tb\) and \(s=S_5/C\), the potential in Eq.~\eqref{eq:sourcepot} becomes
\begin{equation}
 g_5=3(x^4+x^2+\lambda)+\frac{q^2}{x^2}-ts-z\lambda.
 \label{eq:five-source-bare}
\end{equation}
Here both the energy and entropy are affine in the coupling at fixed radius. The common positive-temperature radial domain follows from the coupling-independent numerator in Eq.~\eqref{eq:t5}. At each radius, a coupling endpoint therefore attains the lowest value. Minimizing radially at both endpoints gives the lowest value over the entire interval. For either endpoint coupling \(j\), let \(h_j(t)\) be the radial minimum without the source term. The two endpoint potentials are equal at
\begin{equation}
 z_{\rm coex}^{\rm bare}(t)=\frac{h_{\lambda_U}(t)-h_0(t)}{\lambda_U}.
 \label{eq:five-source-crossing}
\end{equation}
Above this source the upper coupling is preferred, and below it the Einstein endpoint is preferred. This endpoint reduction uses the same bare-length conventions as the four-dimensional calculation. In four dimensions the strictly convex dependence instead allows interior minima.

To implement the common proper bath of Ref.~\cite{across}, write \(t_{\rm cyl}=R T_{\mathrm{b,cyl}}\) and \(\beta_j=\beta(j)\) from Eq.~\eqref{eq:clock-conversion}. The dimensionless bare-clock bath temperature used to find the radius on slice \(j\) is \(t_{\rm cyl}/\beta_j\). After radial minimization the cylinder canonical free energy in units of \(C/R\) is
\begin{equation}
 h_j^{\rm cyl}(t_{\rm cyl})=\beta_j h_j(t_{\rm cyl}/\beta_j)+f_j,
 \label{eq:five-source-cyl}
\end{equation}
Here \(f_j\) is the dimensionless vacuum energy \(f(j)\) in the cylinder convention. Background subtraction (BS) sets \(f=0\). Holographic renormalization (HR) can retain the vacuum energy, as in the Casimir energy of the CFT on the boundary of global Einstein AdS \cite{balasubramanianKraus}. The holographic prescription of Ref.~\cite{across} sets \(f=3\beta^3(3\eta-2)/4\), with \(\eta=\sqrt{1-4\lambda}\).

The cylinder calculation also reduces to the endpoints. At fixed \(x\), put \(A(x)=3(x^4+x^2)+q^2/x^2\). The reduced cylinder energy is \(\beta(A+3\lambda)+f\), and the entropy and source terms are affine in \(\lambda\). The coupling curvature of the full potential therefore equals the energy curvature,
\begin{equation}
 \beta''(A+3\lambda)+6\beta'+f'' .
 \label{eq:cylinder-coupling-curvature}
\end{equation}
Both \(\beta'\) and \(\beta''\) are negative. With background subtraction the curvature is strictly negative. In the holographic prescription \(6\beta'+f''=-3/(16\eta\beta)<0\) \cite{across}, giving strict concavity there as well. The common radial domain then ensures that an interior coupling cannot lower the endpoint minimum.

Let \(J_{\rm cyl}\) be the external source conjugate to \(\lambda\) in this energy convention and \(z_{\rm cyl}=RJ_{\rm cyl}/C\) its reduced value. The endpoint potential is \(h_j^{\rm cyl}-z_{\rm cyl}j\). Equating the endpoint potentials at common \(t_{\rm cyl}\) and \(z_{\rm cyl}\) gives
\begin{equation}
 z_{\rm cyl,coex}=\frac{h_{\lambda_U}^{\rm cyl}-h_0^{\rm cyl}}{\lambda_U}.
 \label{eq:five-source-cyl-crossing}
\end{equation}
Direct cylinder minimization and Eq.~\eqref{eq:five-source-cyl} give the same radii and canonical values.

\begin{figure}[!htb]
\centering
\includegraphics[width=\linewidth]{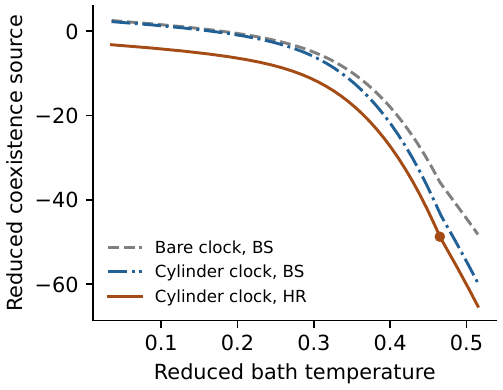}
\caption{Five-dimensional fixed-source coexistence between the bounds \([0,0.09]\) at \(q=0.05\). The axes show the reduced bath temperature in each clock and the reduced coexistence source from Eqs.~\eqref{eq:five-source-crossing} and \eqref{eq:five-source-cyl-crossing}. BS denotes background subtraction and HR denotes holographic renormalization, as in Eq.~\eqref{eq:five-source-cyl}. At each temperature, a source above the corresponding curve favors the upper-bound state. Below it the zero-coupling state is preferred. At the marked point, two black holes at zero coupling and one at the upper bound have equal source potentials.}
\label{fig:five-source}
\end{figure}

Figure~\ref{fig:five-source} compares the curves at \(q=0.05\) and \(\lambda_U=0.09\). On each smooth segment, the slope is minus the entropy difference of the two preferred endpoint states divided by \(C\lambda_U\). In the displayed range the upper endpoint has greater entropy, so heating lowers the source required to make it competitive. For the same numerical value of the dimensionless bath temperature, the two conventions can select different radii. At \(t=0.25\) the upper-endpoint radius is about \(0.42354\). At \(t_{\rm cyl}=0.25\) it is about \(0.46284\). At the same cylinder temperature a change of vacuum energy leaves both radii fixed, and shifts the coexistence source by \((f_{\lambda_U}-f_0)/\lambda_U\simeq-5.48728\). The cylinder calculation also recovers the three-saddle point of Ref.~\cite{across} at \(t_{\rm cyl}\simeq0.465065\) and \(z_{\rm cyl}\simeq-48.71744\) in its holographic scheme. The marked point lies where small and large black holes at the Einstein endpoint already coexist and the source also brings the upper-endpoint state to the same potential. The vacuum term shifts the balance between couplings while remaining constant for the radial competition within a slice.

\section*{Statements and Declarations}

\subsection*{Data and code availability}
The version~1.0.0 reproducibility archive is available on Zenodo~\cite{zengReproducibility}. The record DOI is \href{https://doi.org/10.5281/zenodo.24675964}{10.5281/zenodo.24675964}. It contains the analysis and plotting scripts, numerical tables and residuals, and the archived figure PDFs. The \texttt{run\_all.py} entry point rebuilds the archived plots. Its \texttt{-{}-recalculate} option reruns the symbolic checks and reconstructs the base numerical tables. The manuscript source includes a patch and reproduction instructions for the updated figure labels and layouts.

\subsection*{Use of AI-assisted tools}
OpenAI Codex assisted with drafting, language editing, and implementation of parts of the numerical and plotting workflow. The analytic checks, input data, and verification scripts accompany the manuscript.


\bibliographystyle{spphys}
\bibliography{references}

\end{document}